\documentclass[iop]{emulateapj}
\usepackage[noSeparatorLine]{fltpage}
\usepackage{natbib}
\def\blender{{\tt BLENDER}}
\def\kepler{{\slshape Kepler}}
\def\spitzer{{\it Spitzer}}

\begin{document}

\title{Exoplanet Characterization by Proxy: a Transiting 2.15 $R_{\oplus}$ Planet Near the Habitable Zone of the Late K dwarf Kepler-61}

\author{Sarah~Ballard\altaffilmark{1,2}, David~Charbonneau\altaffilmark{1}, Francois Fressin\altaffilmark{1}, Guillermo~Torres\altaffilmark{1}, Jonathan Irwin\altaffilmark{1}, Jean-Michel~Desert\altaffilmark{3}, Elisabeth~Newton\altaffilmark{1}, Andrew W. Mann\altaffilmark{4}, David~R.~Ciardi\altaffilmark{5}, Justin~R.~Crepp\altaffilmark{3,6}, Christopher E. Henze\altaffilmark{7}, Stephen~T.~Bryson\altaffilmark{7}, Steven~B.~Howell\altaffilmark{7}, Elliott P. Horch\altaffilmark{8}, Mark E. Everett\altaffilmark{9}, Avi Shporer\altaffilmark{10,11,3}}

\altaffiltext{1}{University of Washington, Seattle, WA 98195, USA; sarahba@uw.edu}
\altaffiltext{2}{NASA Carl Sagan Fellow}
\altaffiltext{3}{California Institute of Technology, Pasadena, CA 91125 USA}
\altaffiltext{4}{Institute for Astronomy, University of Hawai'i, Honolulu, HI 96822}
\altaffiltext{5}{NASA Exoplanet Science Institute/Caltech, Pasadena, CA 91125}
\altaffiltext{6}{Department of Physics, University of Notre Dame, Notre Dame, IN 46556}
\altaffiltext{7}{NASA Ames Research Center, Moffett Field, CA 94035, USA}
\altaffiltext{8}{Southern Connecticut State University, New Haven, CT 06515}
\altaffiltext{9}{National Optical Astronomy Observatory, Tucson, AZ 85719}
\altaffiltext{10}{Las Cumbres Observatory Global Telescope Network, Santa Barbara, CA 93117}
\altaffiltext{11}{Department of Physics, University of California, Santa Barbara, CA 93106}


\begin{abstract}
We present the validation and characterization of Kepler-61b: a 2.15 $R_{\oplus}$ planet orbiting near the inner edge of the habitable zone of a low-mass star. Our characterization of the host star Kepler-61 is based upon a comparison with the set of spectroscopically similar stars with directly-measured radii and temperatures. We apply a stellar prior drawn from the weighted mean of these properties, in tandem with the \kepler\ photometry, to infer a planetary radius for Kepler-61b of 2.15$\pm$0.13 $R_{\oplus}$ and an equilibrium temperature of 273$\pm$13 K (given its period of 59.87756$\pm$0.00020 days and assuming a planetary albedo of 0.3). The technique of leveraging the physical properties of nearby ``proxy'' stars allows for an independent check on stellar characterization via the traditional measurements with stellar spectra and evolutionary models. In this case, such a check had implications for the putative habitability of Kepler-61b: the planet is 10\% warmer and larger than inferred from $K$--band spectral characterization. From the \kepler\ photometry, we estimate a stellar rotation period of 36 days, which implies a stellar age of $>1$ Gyr. We summarize the evidence for the planetary nature of the Kepler-61 transit signal, which we conclude is 30,000 times more likely to be due to a planet than a blend scenario. Finally, we discuss possible compositions for Kepler-61b with a comparison to theoretical models as well as to known exoplanets with similar radii and dynamically measured masses. 
\end{abstract}

\section{Introduction}
With the discoveries of exoplanets Kepler-22b \citep{Borucki11}, Kepler-20e \& f \citep{Fressin12}, Kepler-42a, b, \& c \citep{Muirhead12a}, Kepler-68c \citep{Gilliland13}, and Kepler-62e \& f \citep{Borucki13}, astronomers are encroaching upon the regime of transiting terrestrial exoplanets in their stellar habitable zones. Kepler-22b is the first super-Earth-sized exoplanet with a measured radius to reside in the habitable zone of a sun-like star, though its radius of 2.4 $R_{\oplus}$ does not necessitate a terrestrial composition. The Kepler-20, Kepler-42, and Kepler-68 exoplanetary systems each comprise multiple planets, some of which are Earth-sized or smaller (as small as Mars in the case of Kepler-42c). However, these planets orbit too close to their host stars to lie within the habitable zone. The star Kepler-62 hosts five planets, two of which are both very likely solid and reside in their star's habitable zone \citep{Borucki13}. The most recent release of \kepler\ exoplanetary candidates \citep{Batalha13} contains 10 members $<$2 $R_{\oplus}$ and with equilibrium temperatures between 185 and 303 K. This temperature range is a generous definition of the habitable zone proposed by \cite{Kasting12}. Half of these candidate exoplanets orbit stars cooler than 4100 K, as reported by the \kepler\ Input Catalog.

However, inferring the properties of low-mass stars from spectra (upon which a measurement of planetary radius and equilibrium temperature hinges so critically) presents difficulties on multiple fronts. The direct comparison of theoretical spectra to observations, which is robust for deducing the properties of solar-type stars, is challenging for low-mass stars. Such spectra rely on detailed, computationally intensive modeling of convection in low-mass stellar interiors \citep{Mullan01, Browning08} and complete lists of the complex array of molecules and grains that reside in their atmospheres \citep{Tsuji96, Allard00}. For this reason, we often appeal to empirical, rather than theoretical, methods for the physical characterization of low-mass stars (see \citealt{Torres11b} for a complete review). This challenge is compounded by the possibility that stellar properties may also depend on other parameters, such as activity and metallicity, in a significant way. The empirical technique for deriving M dwarf temperatures and metallicities from $K$-band spectra that was innovated by \cite{Rojasayala12} offers an important inroad. However, this technique depends upon the H$_{2}$O-K2 spectral index, which is effective probe of stellar temperature for mid-M dwarfs, but saturates for stars with temperatures higher than 3900 K \citep{Muirhead12b}.  There exists a desert in stellar temperature, near 4000 K, where no reliable method method for temperature derivation from a spectrum exists: the H$_{2}$O-K2 index has saturated, and the star is yet too cool for comparison between high-resolution optical spectra and synthetic models. This is an especially salient problem for the characterization of the current and future sample of exoplanets orbiting low-mass stars, given the astonishing occurrence rates of 1.0 planet/star (found from the \kepler\ sample to be 0.90$^{+0.04}_{-0.03}$ planets/star per \citealt{Dressing13} and 1.0$\pm$0.1 planets/star per \citealt{Swift13}).

There exists an alternative means of measuring the properties of the nearest and brightest low-mass stars, with measured distances from parallax: interferometric measurements of their radii. The radius measurement, in tandem with the bolometric flux, also enables a direct measurement of the stellar temperature with minimal modeling uncertainties. The number of low-mass stars with directly measured properties from interferometry is growing, and this pool of stars can be plumbed for proxies to stars too faint for such direct characterization themselves. \cite{Muirhead12a} undertook the first such analysis with an application of the properties of Barnard's star toward a characterization of the M4V star and three transiting planets comprising the Kepler-42 system. A similar method would be especially useful for a star astride the 4000 K boundary, where spectroscopic estimates of the stellar properties may be unreliable, and for an exoplanetary host star. Kepler-61 is such a star: its temperature lies near to 4000 K, and depending upon the source of its stellar characterization, the equilibrium temperature of its planet lies either outside or astride the stellar habitable zone. The radius of the planet varies from 2.0 to 2.3 $R_{\oplus}$ (\citealt{Muirhead12a} and \citealt{Burke13}, respectively) depending on the assumed size of the star, which range brackets both a plausible rocky composition (more amenable to habitability) or a ``mini-Neptune'' composition. Moreover, the now exists a sample of four stars with (a) similar spectral type and (b) direct radius and temperature measurements, which can be applied to break the degeneracy of the planet's putative habitability.

The apparent magnitude of Kepler-61, with \kepler\ magnitude Kp=15.0, renders the star too dim to enable a radial velocity measurement of the planet's mass. However, even without a mass measurement of Kepler-61b, we are able to validate its authentic planetary nature with a statistical argument about the likelihood of the planet scenario in comparison to false-positive scenarios. We undertake such an analysis with \blender, which machinery has already been applied to validate Kepler-9d \citep{Torres11a}, Kepler-11g \citep{Lissauer11}, Kepler-10c \citep{Fressin11}, Kepler-19b \citep{Ballard11b}, Kepler-22b \citep{Borucki12}, and Kepler-20 e \& f \citep{Fressin12}. In this case, a single observation of the transit depth at 4.5 $\mu$m with Warm \spitzer\, in tandem with the \blender\ result, plays a prominent role in ruling out hierarchical triple false-positive scenarios.


In Section 2, we describe the \kepler\ observations of Kepler-61. In Section 3 we describe our characterization of the transit light curve and the physical parameters of the star, including our method of applying the properties of nearby similar stars to characterize Kepler-61. In Section 4, we describe the validation of Kepler-61b as an authentic planet with \blender. We include a description of follow-up imaging observations of the star to characterize any additional stars within the \kepler\ photometric aperture of Kepler-61, as measurements of the photocentroid gather from the \kepler\ images and a measurement of the transit depth with Warm \spitzer. In Section 5, we comment on the stellar rotation of Kepler-61 and transit times of Kepler-61b. We also discuss plausible compositions for the planet, given the growing number of transiting super-Earth-sized planets with dynamically-measured masses. Finally, in Section 6, we describe future prospects. 

\section{{\slshape Kepler} Observations}
The {\slshape Kepler} spacecraft, launched on 7 March 2009, is photometrically monitoring 170,000 stars for 8 years for evidence of transiting planets. \cite{Argabright08} provides an overview of the {\slshape Kepler} instrument, and \cite{Caldwell10} and \cite{Jenkins10} provide a summary of its performance since launch. The {\slshape Kepler} observations of Kepler-61 (\kepler\ Input Catalog number 6960913) that we present in this work were gathered from 13 May 2009 to 3 October 2012, spanning \kepler\ ``Quarters'' 1--14. All data for this star were gathered in long-cadence mode (characterized by an exposure time of 29.5 minutes) for Quarters 1-11, and in short-cadence mode (characterized by an exposure time of 58.5 s) for Quarters 12, 13, and 14. The data contain gaps of approximately 3 days between quarters for scheduled downlinks. Kepler-61b was first identified as exoplanetary candidate Kepler Object of Interest (KOI) 1361.01 by \cite{Borucki11}. We employed the light curves generated by the \kepler\ aperture photometry (PDC-Map) pipeline, described in \cite{Twicken10}, to which we add an additional step. We remove the effects of baseline drift by individually normalizing each transit as follows. We fit a linear function of time to the flux immediately before and after transit (specifically, from 9 hours to 20 minutes before first contact, equal to 2.5 transit durations, and an equivalent time after fourth contact). 

\section{Analysis}

\subsection{Derivation of Stellar Parameters}
While the physical characterization of isolated low-mass stars is a notoriously difficult problem \citep{Segransan03, Torres11b}, several recent techniques have offered promising inroads by tying spectra of M dwarfs to directly measured quantities. Such characterization of low-mass stars is crucial, as the {\it Kepler} mission has demonstrated that low-mass stars are hosts to small planets at a rate of 1.0 planet/star \citep{Dressing13,Swift13}. 


\subsubsection{Characterization in the Literature}
Kepler-61 is classified as an M0 star by \cite{Muirhead12b}, who employed $K$-band spectra of the star to infer stellar properties (in this work, they refer to Kepler-61b by its KOI notation, KOI 1361.01). They measure ratios of equivalent widths of Na and Ca to determine the stellar metallicity, and deformation between continuum regions within $K$--band  (the H$_{2}$O-K2 index, first developed by \citealt{Covey07} and re-calibrated by \citealt{Rojasayala12}), which they interpolate onto the theoretical metallicity and H$_{2}$O-K2 surface from the BT-Settl late-type model spectra \citep{Allard12} to determine the stellar effective temperature. The metallicity relation published by \cite{Rojasayala12} is calibrated using binaries comprising an M dwarf and an FGK star, which \cite{Muirhead12b} then applied to a set of isolated low-mass stars in the \kepler\ sample.  For Kepler-61, they find $T_{\rm eff}$ of 3929$^{+66}_{-135}$ K and metallicity [Fe/H] of -0.02$\pm$0.11 (though they caution that applying the $K$-band metallicity metric to stars with temperatures higher than 3900 K relies upon an extrapolation of the \citealt{Rojasayala12} metric). By comparing this temperature and metallicity to the Dartmouth stellar evolutionary models \citep{Dotter08}, they determine a stellar radius $R_{\star}$= 0.55$\pm$0.07 $R_{\odot}$ and a mass $M_{\star}$= 0.57$\pm$0.07 $R_{\odot}$. However, as described above, this location in temperature space is close to the location where the H$_{2}$O-K2 index saturates (for $T_{\mbox{eff}}>3900$ K, \citealt{Muirhead12b}), and the deformation between continuum regions is too small to effectively probe stellar temperature. \cite{Dressing13} corroborates the result that temperatures derived from the  H$_{2}$O-K2 index are significantly lower than those derived from the comparison of the broadband colors to models for stars near the 3900 K marker. The \cite{Dressing13} method relies upon a comparison of the measured magnitudes of the star from the \kepler\ Input Catalog (2MASS $JHK$ and Sloan filters $g$, $r$, $i$, and $z$) against the colors predicted from the Dartmouth stellar evolutionary models \citep{Dotter08}. They assign a prior on stellar metallicity based upon the metallicity distribution of the M dwarfs observed in the \cite{Casagrande08} sample and a prior on height above the galactic midplane similar to that applied by \cite{Brown01} for the Kepler Input Catalog. For the colors of Kepler-61, they estimate an effective temperature of 4060$^{+100K}_{-109}$ K, a radius of 0.57$^{+0.06}_{-0.11}$ $R_{\odot}$, and a mass of 0.57$^{+0.08}_{-0.09}$ $M_{\odot}$ (C. Dressing, private communication).

\subsubsection{Optical and Near-Infrared Spectroscopy}
\begin{figure}
\begin{center}
 \includegraphics[width=5in]{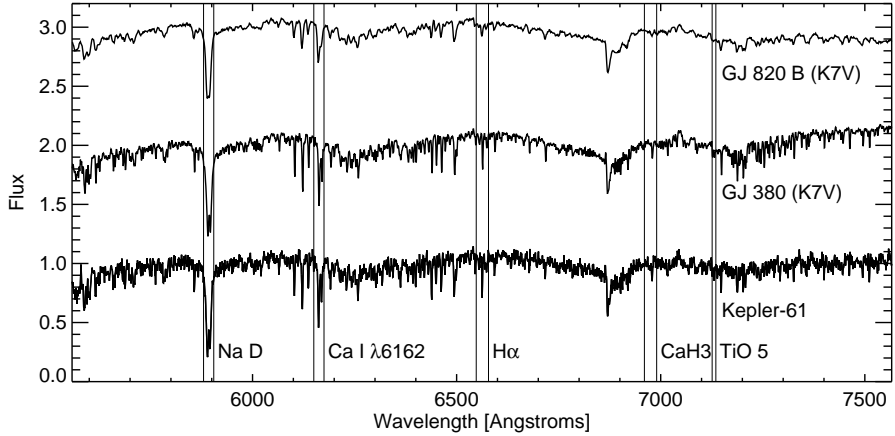} 
 \caption{FAST spectra of GJ~820B (top), GJ~380 (center) and Kepler-61 (bottom).  We have denoted the wavelength regions employed by spectral typing software, ``The Hammer'', developed by \cite{Covey07}.}
  \label{fig:FAST}
\end{center}
\end{figure}

Using the FAST spectrograph on the 1.5 m telescope at Mount Hopkins, AZ, we gathered a spectrum of Kepler-61 in the range 5560--7570 $\AA$ with 0.75 $\AA$ resolution (we employed an integration time of 20 minutes to achieve a signal-to-noise ratio of 30). In Figure \ref{fig:FAST}, we show the Kepler-61 spectrum in comparison to spectra of two nearby K7V stars, GJ~380 and GJ~820B, gathered with the same instrument (we observed the former on 24 April 2012 with integration time 10 s for a signal-to-noise ratio of 210 pixel$^{-1}$ and gathered the latter spectrum from the FAST Spectrograph Archive. It was observed on 4 November 2010 at resolution of 1.5$\AA$ and an integration time of 2 s for a signal-to-noise ratio of 200 pixel$^{-1}$). We have denoted the wavelength regions employed by spectral typing software, ``The Hammer'', developed by \cite{Covey07}, which we use to determine a spectral type of K7V for the KOI. \cite{Lepine13} find that classification with the Hammer agrees with stellar classification from spectral indices within 1.0 subtypes.

We consider the set of similarly typed stars with directly measured radii. While spectral types for stars in the M0-K7 range compiled in the literature often vary by 1-2 subtypes, we defer to the spectral types listed in \cite{Boyajian12}.  In this case, because the \cite{Dressing13} predicted temperature for Kepler-61 encompasses an effective temperature as high as 4160 K (within 1$\sigma$), we elect to exclude K5 stars from the sample of spectroscopically similar stars with resolved radii: the K5 stars GJ~820A and GJ~720B have effective temperatures of 4361$\pm$17 \citep{Kervella08, vanBelle09, Boyajian12} and 4393$\pm$149 \citep{Boyajian12} respectively, which lowest temperature estimates within 1$\sigma$ are still hotter than the 1$\sigma$ range predicted for temperature of Kepler-61 from broadband photometry. The sample of K7 and M0 stars with radius measurements currently comprises four stars: GJ~380 (radius and temperatures measured gathered by \citealt{Lane01}, \citealt{vanBelle09}, and \citealt{Boyajian12}), GJ~338A (measurements from \citealt{Boyajian12}), GJ~338B (measurements from \citealt{Boyajian12}), and GJ~820B (measurements from \citealt{Kervella08} and \citealt{vanBelle09}). We list the properties of these stars in Table 1. 

\begin{figure}[h!]
\begin{center}
 \includegraphics{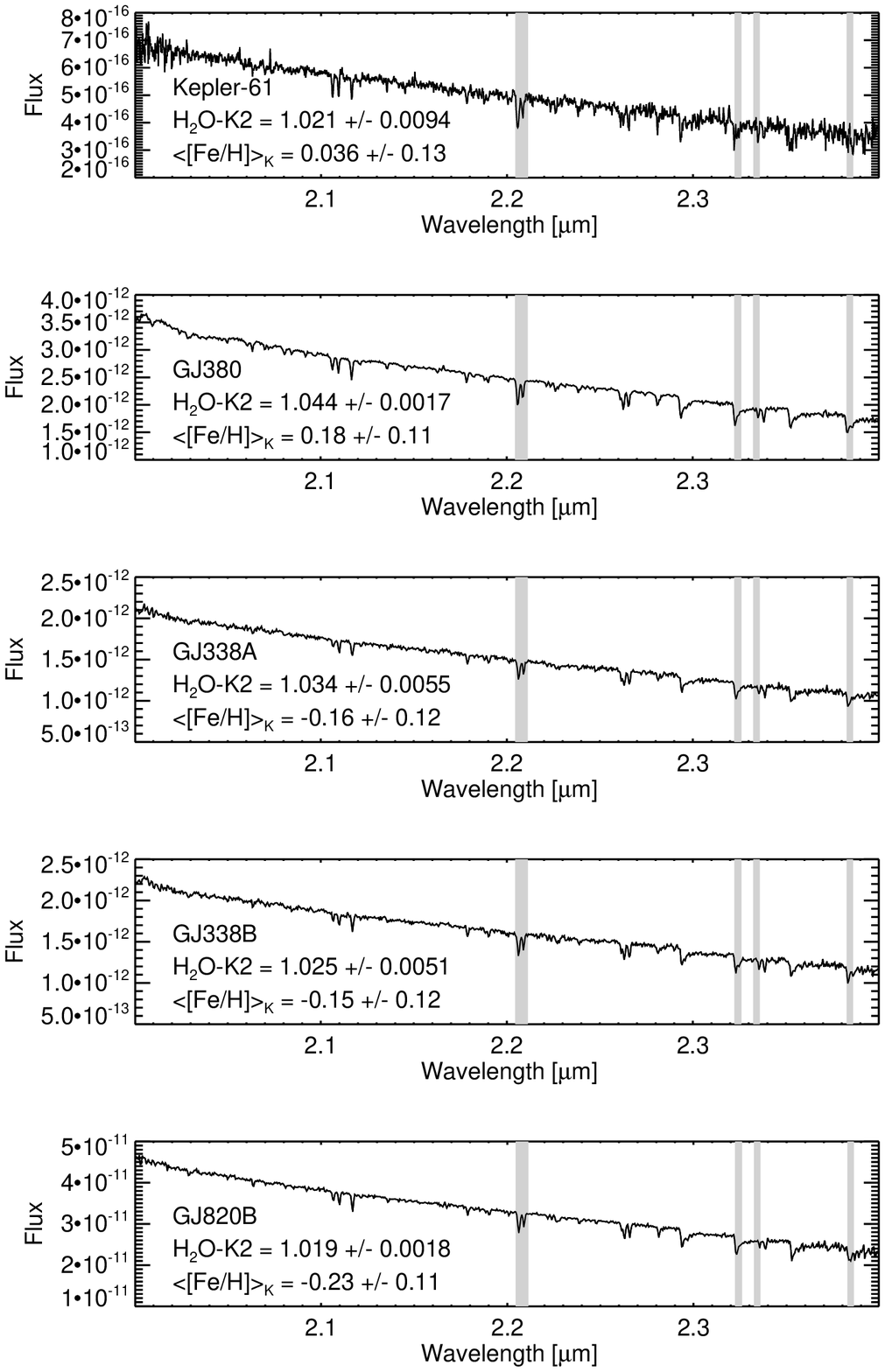} 
 \caption{$K$--band spectra for Kepler-61 (KOI 1361) and for nearby stars of similar spectral type. Overplotted in gray are the metal-sensitive regions published by \cite{Mann12}; we use the metric described in that work to calculate [Fe/H] for each star. The H$_{2}$O-K2 temperature index \citep{Rojasayala12} is also stated for each star.} 
  \label{fig:kband}
\end{center}
\end{figure}

We have compiled the set of $K$-band spectra for these stars in addition to the KOI, which we depict in Figure \ref{fig:kband}, in order to estimate their [Fe/H] metallicities in a uniform fashion from the metric of \cite{Mann12} (which metric is valid for stars as early as spectral type K5). Our spectrum for Kepler-61 was gathered and published by \cite{Muirhead12a} with the TripleSpec instrument at Palomar Observatory \citep{Herter08}. They employed an exposure time of 6 minutes on 6 June 2011, with resolution of 3$\AA$ ($R$ of 2700), to obtain a signal-to-noise ratio of 60 pixel$^{-1}$. We gathered spectra for GJ~338A and GJ~338B on 27 Jan 2013 with the SpeX instrument at NASA's Infrared Telescope Facility (IRTF, \citealt{Rayner03}), and for GJ~380 on 17 December 2012.  We employed the ShortXD observing mode (resolution of 5$\AA$, $R$ of 2000) and exposure times of 1s, 1s, and 5s, respectively, for signal-to-noise ratios of 150, 150, and 700 pixel$^{-1}$. We gathered our spectrum of GJ~820B from the IRTF Spectral Library \citep{Rayner09}. It was observed on 2001 October 20 with the same resolution, and has a signal-to-noise ratio of 800 pixel$^{-1}$. We compute uncertainties on these metallicities from adding the scatter in the $K$-band metric quoted by \cite{Mann12} of 0.11 dex in quadrature to the intrinsic uncertainty in the value of [Fe/H]$_{K}$ from the error in the spectrum at the wavelengths that are operative for the metric. We list these derived metallicities in Table 1, but note that no standard star possesses a metallicity consistent with 1$\sigma$ of the most probable [Fe/H] value for Kepler-61 of 0.03. Unlike in the case of Kepler-42 \citep{Muirhead12b}, none of these nearby stars possesses features consistent enough with the KOI to render one of them an single ideal ``proxy'' star for the Kepler target.

\subsubsection{Activity and Age Indicators}
\begin{figure}[h!]
\begin{center}
 \includegraphics[width=5in]{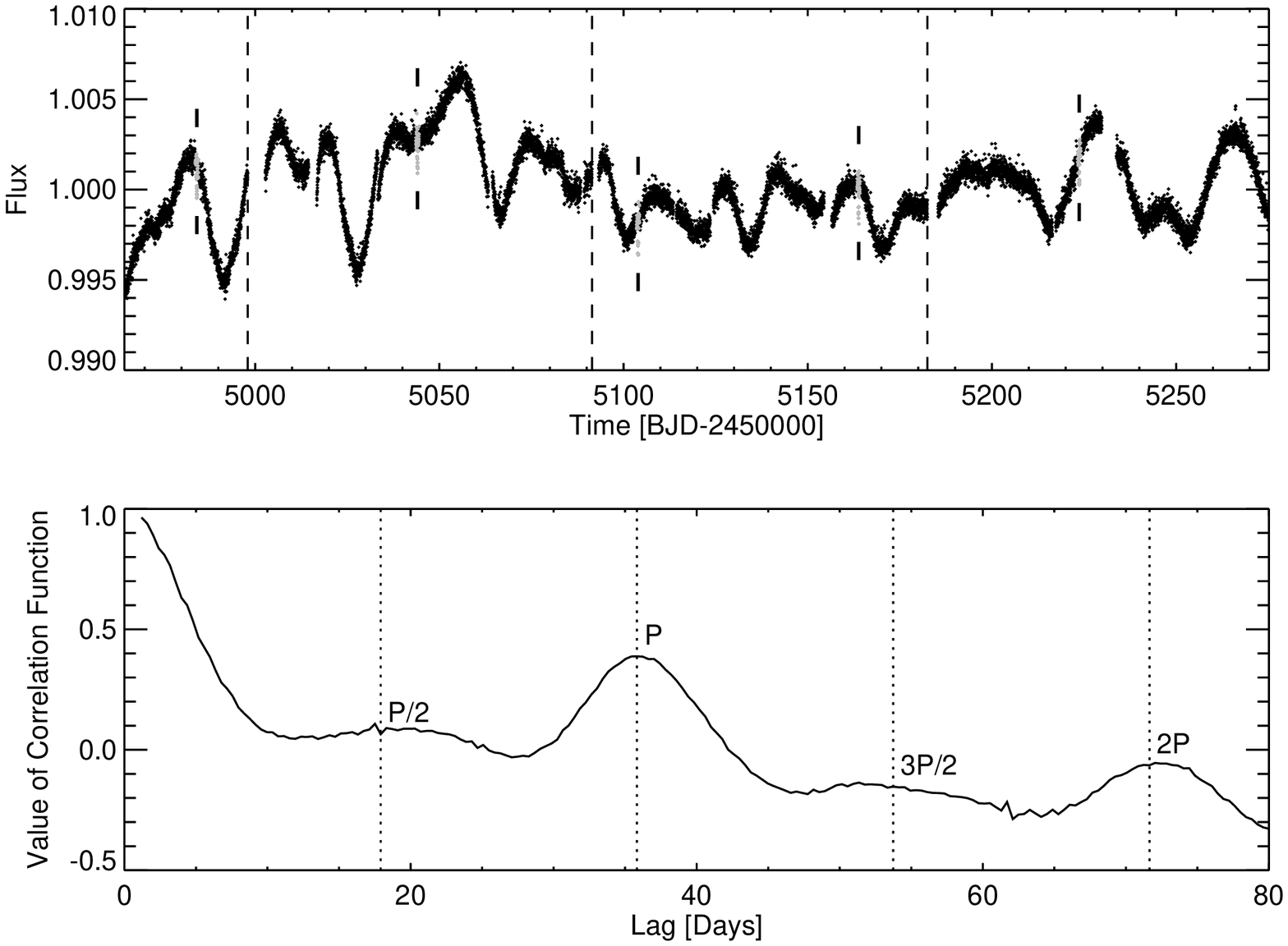} 
 \caption{{\it Top:} Quarters 1--4 of \kepler\ observations of Kepler-61. Dashed lines denote the intervals between quarters, and observations in transit are indicated in gray, with vertical lines above and below transit times. {\it Bottom:} The results of the discrete correlation function applied to this portion of the light curve. The strongest peak corresponds to 36 days, which variability is present by eye in the 300 days of observations depicted in the top panel. Dotted lines indicate the strongest period and its harmonics.}
  \label{fig:rotation}
\end{center}
\end{figure}

We consider the activity levels of the stars in this sample, in comparison to Kepler-61. We have measured the equivalent widths in H${\alpha}$ for both Kepler-61 from the FAST spectra to those compiled by \cite{Gizis02}. They exhibit equivalent widths in H$\alpha$ (all in absorption) ranging from -0.82 (for Kepler-61) to -0.50 (for GJ~338A). In addition, we have measured the rotation period for Kepler-61 from the \kepler\ photometry, which we compare to the rotation periods listed for three of the four nearby stars in \cite{Barnes07}. The ``gyrochronology'' technique of mapping the rotational period of a star to its age is described in detail in \cite{Barnes03,Barnes07}, and is used specifically in \cite{Barnes07} to estimate ages of 1.96 Myr, 1.36 Myr, and 2.96 Gyr for GJ 380, GJ~338A, and GJ~820, respectively. In Figure \ref{fig:rotation}, we depict the first four quarters of \kepler\ observations of Kepler-61, or approximately one year of continuous observation. These observations have been processed using the Presearch Data Conditioning (PDC) module of the Kepler data analysis pipeline, with the Bayesian Maximum A Posteriori (MAP) approach applied (described in \citealt{Smith12}). The use of highly correlated and quiet stars to create a set of co-trending basis vectors enables the removal of non-astrophysical artifacts from the \kepler\ time series, and the preservation of astrophysically interesting signals such as stellar rotation. We apply the discrete correlation function of \cite{Edelson88}, similarly applied by \cite{Fabrycky12a} on the time series of Kepler-30 and \cite{Queloz09} on CoRoT-7, on this portion of the Kepler-61 light curve. We test lags from 1 to 100 days,  and identify a 36$\pm$4 day periodicity, which we attribute to the stellar rotation period. In Figure \ref{fig:rotation}, we show both the \kepler\ photometry and autocorrelation function that we employed to characterize the stellar rotation. \cite{Irwin11} recently published a compilation of the known rotation periods of low-mass stars from the literature. The rotation periods are drawn from open clusters of stars with derived ages from 1--650 Myr (which ages are measured by main sequence fitting of these clusters, as compared to stellar evolutionary models), and then also from field stars with ages $>$1 Gyr. Among the clusters with ages less than 1 Gyr, nearly all of the stars with masses $>$0.5 $M_{\odot}$ have rotation periods shorter than 30 days, and only after 1 Gyr do low-mass stars with masses greater than 0.5 $M_{\odot}$ appear to spin down enough to produce 36 day rotation periods. Stars with masses between 0.5 and 0.7 $M_{\odot}$ and ages between 8 and 10 Gyr are more likely to have rotation periods in the tens of days (\citealt{Kiraga07} and \citealt{Baliunas96} observed values $<$30 days for field stars in the 1-2 Gyr range). We therefore take the observed rotation period of Kepler-61 to be conservatively indicative of an age $>$1 Gyr.

 We also apply the age metric of \cite{Barnes10} to estimate the age of Kepler-61, which \cite{Swift13} used to obtain an age approximation for Kepler-32:

\begin{equation} 
t=\frac{\tau}{k_{c}}\mbox{ln}\left( \frac{P}{P_{0}} \right)+\frac{k_{I}}{2\tau}(P^{2}-P_{0}^{2}),
\end{equation}

\noindent where the dimensionless constants $k_{C}$ = 0.646 days Myr$^{−1}$ and $k_{I}$ = 452 Myr day$^{−1}$ are approximated in \cite{Barnes10}. We estimate the convective turnover time $\tau$ from \cite{Wright11}, who derived empirical convective turnover times for a sample of 824 solar and late-type stars, the typical convective turnover time of a star within the mass range 0.47--0.62 $M_{\odot}$ is 29 days. It's not possible to determine the initial spin period $P_{0}$ associated with Kepler-61 at its birth, so we defer to the median value of the initial spin period, $P_{0}$ = 2.81 days required to produce the observed rotation rates for 0.6 $M_{\odot}$ stars in the Praesepe cluster \citep{Agueros11} similarly to \cite{Swift13}. This formulation returns an age for Kepler-61 of 10 Gyr. We therefore find that ages between 1-10 Gyr are consistent with different metrics of age constraints for Kepler-61, and simply adopt a lower bound on its age of 1 Gyr. 

Because none of these nearby stars comprises an ideal ``proxy'' to the planet-host star (i.e., possessing both statistically indistinguishable metallicities and temperature indices in tandem with similar activity indicators), we adopt the conservative tack of employing a radius and temperature for Kepler-61 which are the weighted mean of the radii and temperatures of the set of standard stars. For our uncertainty on these values, we encompass the highest and lowest mean value among the sample. We therefore adopt for Kepler-61 a radius of 0.62$^{+0.02}_{-0.05}$ $R_{\odot}$ and a temperature of 4017 $^{+68}_{-150}$ K. 


\begin{deluxetable*}{rrrrrr}
\tabletypesize{\scriptsize}
\singlespace
\tablecaption{Comparison of Observables between Kepler-61 and Similar Stars}
\label{tbl:spex}
\tablewidth{0pt}
\tablehead{
\colhead{Parameter} & \colhead{GJ~380} & \colhead{GJ~338A} & \colhead{GJ~338B} & \colhead{GJ~820B} & \colhead{Kepler-61}\\}
\startdata
\hline
Spectral Type\tablenotemark{a} & K7V & M0V  & K7V & K7V & K7V \\
Metallicity [Fe/H] & 0.18$\pm$0.11 & -0.15$\pm$0.12 & -0.15$\pm$0.12 & -0.23$\pm$0.11 & 0.03$\pm$0.14 \\
H$_{2}$O-K2 & 1.044$\pm$0.002 & 1.034$\pm$0.005 & 1.025$\pm$0.005 & 1.019$\pm$0.002 & 1.02 $\pm$0.010\\
E$_{\mbox{H}\alpha}$\tablenotemark{b} (in absorption)& -0.61 & -0.56 & -0.50 & -0.59 & -0.82 \\
log($L_{X}/L_{bol}$)\tablenotemark{a} &-5.16 & -4.68 & -4.65 & -5.03 & --  \\ 
Rotation period [days]\tablenotemark{c} & 11.67 & -- & 10.17 & 37.9/48 & 36 \\
Estimate Age [Gyr]\tablenotemark{c} & 0.2 & 0.1 & -- & 3 & -- \\
\hline
$R_{\star}$ [$R_{\odot}$]\tablenotemark{a} & 0.64$\pm0.004$ & 0.58$\pm0.01$ & 0.57$\pm0.01$ & 0.60$\pm0.007$ & 0.62$^{+0.02}_{-0.05}$ (adopted) \\ 
$T_{\mbox{eff}}$ [K]\tablenotemark{a} & 4085$\pm14$ & 3907$\pm35$ & 3867$\pm35$ & 3932$\pm25$ & 4017$^{+68}_{-150}$(adopted) \\
\enddata
\tablenotetext{a}{Values for standard stars from \cite{Boyajian12}.}
\tablenotetext{b}{Values for standard stars from \cite{Gizis02}.}
\tablenotetext{c}{From \cite{Barnes07}, stated to nearest 100 Myr.}
\end{deluxetable*}


\begin{figure}
\begin{center}
 \includegraphics[width=5in]{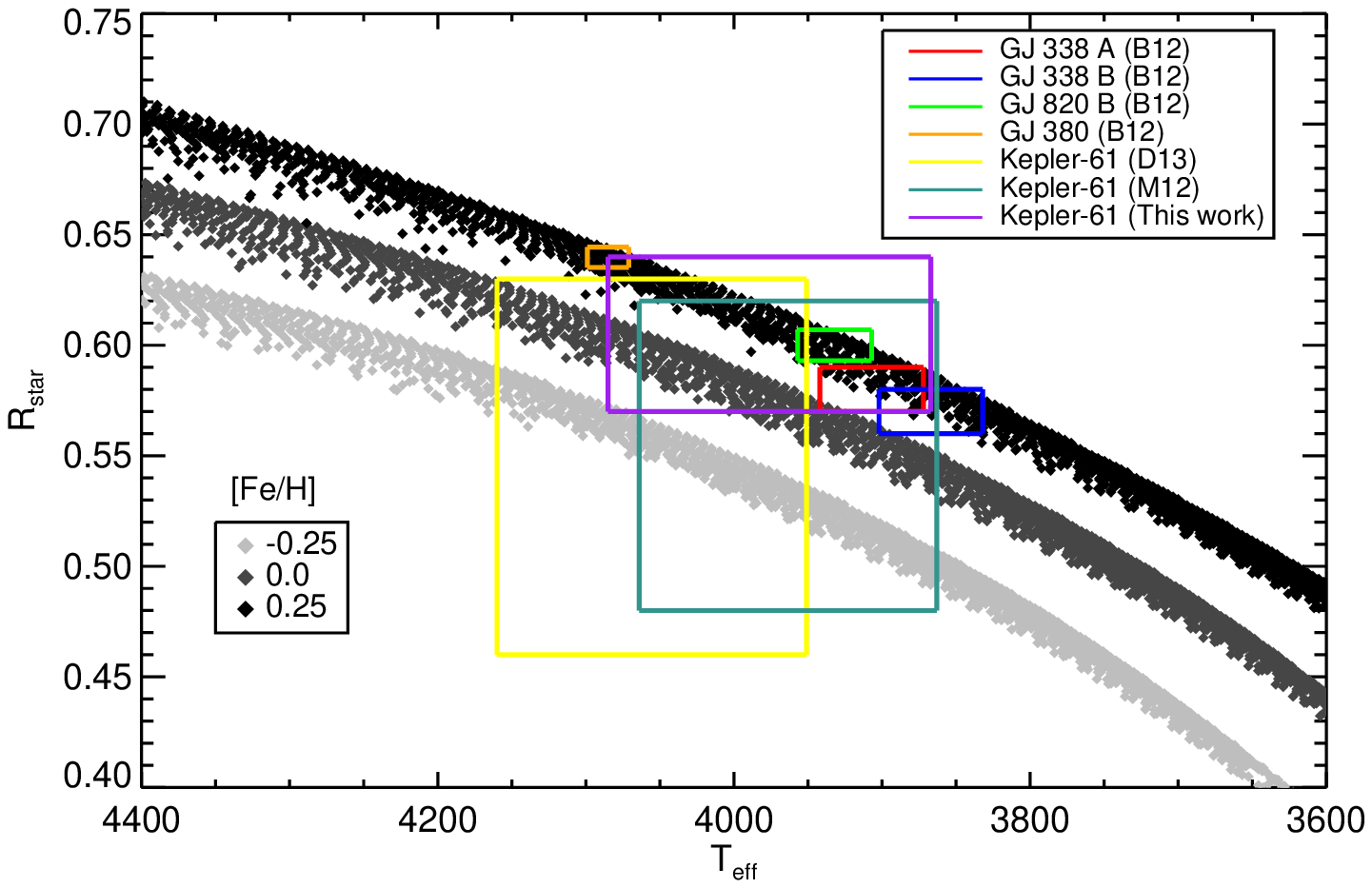} 
 \caption{Stellar effective temperatures versus radius for Kepler-61 and for nearby standard stars, with stellar evolutionary models from Dartmouth \citep{Dotter08} overplotted, from ages ranging from 1--14 Gyr, at values of [Fe/H] equal to -0.25, 0.0, and 0.25. The directly-measured radii and temperatures of GJ~380, GJ~338A \& B, and GJ~820 B are drawn from \cite{Boyajian12}. We employ a radius and effective temperature for Kepler-61 from the weighted mean of these values, with an error bar that encompasses the highest and lowest means in the sample. The estimates for Kepler-61 from $K$--band spectra \citep{Muirhead12a} and from the broadband KIC photometry \citep{Dressing13} are also depicted.}
  \label{fig:perturb}
\end{center}
\end{figure}

\subsection{Derivation of Planetary Parameters}
We estimated the uncertainty in the planetary transit parameters using the Markov Chain Monte Carlo (MCMC) method as follows. We employ model light curves generated with the routines in \cite{Mandel02}, which depend upon the period $P$, the epoch $T_{c}$, the planet-to-star radius ratio $R_{p}/R_{\star}$, the ratio of the semi-major axis to the stellar radius $a/R_{\star}$, the impact parameter $b$, the eccentricity $e$, and the longitude of periastron, $\omega$. We fixed two quadratic limb-darkening coefficients (LDCs), $u_{1}$ and $u_{2}$, to theoretical values based on the adopted effective temperature and radius. We employed the theoretical limb darkening coefficients generated for the \kepler\ bandpass by \cite{Claret11} from the PHOENIX models corresponding to a star with $T_{\rm eff}$ of 4000 K and solar metallicity, which is the closest match to the effective stellar temperature $T_{\rm eff}$ of 4017 and the metallicity of 0.03$\pm$0.14 for Kepler-61 (we additionally specified a log(g) of 4.5 and an intermediate turbulent velocity value of 2 km s$^{-1}$): these coefficients are $u_{1}$=0.50 $u_{2}$=0.20. We accounted for the 29.5 minute integration time of the \kepler\ photometry, which is three times longer than the ingress and egress duration of the planet candidate, by evaluating the light curve model at intervals of 1 minute, and then summing the model over the long cadence integration time. We model the three quarters of short cadence observations independently (gathered from Quarters 12--14, during which time the planet presented five transits). For these short cadence observations, we evaluate the light curve model at each time measurement, in 58.5 s intervals, and fix the period and transit time to the best-value recovered over the 11-quarter baseline of long-cadence observations. 

To generate the MCMC chain, we randomly choose one parameter, perturb it, and evaluate the $\chi^{2}$ of the solution. If the $\chi^{2}$ is lower, we accept the new parameter value. If the $\chi^{2}$ is higher, we evaluate the probability of accepting the jump as $p=e^{-\Delta\chi^{2}/2}$. If the jump is rejected, the procedure is repeated at that point in the chain until an acceptable jump occurs. We adjust the width of the distribution from which we randomly draw the jump sizes in each parameter until 20--40\% of jumps are executed in each of the parameters. We created five chains, each of length 10$^{6}$ points, where each of the chains is begun from a different set of starting parameters (each parameter is assigned a starting position that is +$3\sigma$ or -$3\sigma$ from the best-fit values). We discard the first 20\% of jumps from each chain to remove the transient dependence of the chain on the starting parameters. We first conducted this analysis, as described, using only the \kepler\ light curve to inform our value of $\chi^{2}$. However, the allowable stellar densities we infer from the light curve alone are much broader than the range of stellar densities consistent with our spectroscopic information about the star. Because Kepler-61 is a late K star, it is slowly evolving-- therefore, its range of theoretical densities is tightly constrained for ages $<14$ Gyr. When we apply the MCMC algorithm toward fitting the long-cadence transit parameters independently and allow $a/R_{\star}$ to float, we find that values of $a/R_{\star}$ from 35--150 furnish comparable fits to the light curve (the ingress and egress time, at 9.3$\pm$3.2 minutes measured at long-cadence, results in a wide family of allowable light-curve fits). We take advantage of two circumstances that allow us to better constrain the transit parameters. First, there exist 5 transits gathered by \kepler\ at short-cadence, where the ingress and egress time are resolved by the 1-minute exposure time. Second, we make use of the fact that low-mass stars are slowly evolving to set a physically-motivated prior on $a/R_{\star}$ as follows. We based our procedure for constraining the mass, radius, and age of the host star on the method described by \cite{Torres08}. Using the metallicity we derive from $K$ band, we created a set of stellar evolution models from the Dartmouth isochrone series \citep{Dotter08}. We employed the interpolation software that accompanied that work, which accepts as inputs the age of the star, the iron abundance, and the abundance of $\alpha$-elements relative to solar (for which we assume the solar value), and outputs a grid of stellar isochrones corresponding to a range of masses. We evaluated a set of isochrones over an age range of  1 to 14 Gyr (at intervals of 0.1 Gyr) and in [Fe/H] in increments of 0.01 from -0.5--0.5 dex (encompassing 3$\sigma$ above and below the measured [Fe/H] of 0.03 $\pm$ 0.14). We evaluate the physical radius corresponding to each stellar model via log(g) and the mass of the star ($g=GM_{\star}/R_{\star}^{2}$).

Rearranging Kepler's version of Newton's third law in the manner employed by \cite{Seager03}, \cite{Sozzetti07} and \cite{Torres08}, we convert the period (derived from photometry), and the radius and mass of the host star (from isochrones) to a ratio of the semi-major axis to the radius of the host star, $a/R_{\star}$:

\begin{equation}
\frac{a}{R_{\star}}=\left(\frac{G}{4\pi^{2}}\right)^{1/3}\frac{P^{2/3}}{R_{\star}}(M_{\star}+M_{p})^{1/3},
 \label{eq:kepler_5}
\end{equation}
\noindent where we will hereafter assume that $M_{p}$ is negligible when compared to the mass of the host star. 

We calculate the corresponding value of $a/R_{\star}$ for each stellar model. We then generate the MCMC chain as follows. We implement a prior on $a/R_{\star}$ by varying the adopted quantities $R_{\star}$ and $T_{\mbox{eff}}$ in the chain, in addition to the light curve parameters. For each set of $R_{\star}$ and $T_{\mbox{eff}}$,  we locate the closest stellar model associated with these values and record its corresponding stellar density, $a/R_{\star}$. It is this value of $a/R_{\star}$ that is used to generate the light curve model, along with the other light curve parameters $P$, $T_{c}$, $R_{p}/R_{\star}$, $b$, $e$, and $\omega$ (defined above), which are permitted to vary independently. We assign uniform flat priors in $b$ (from 0--1), $e$ (from 0--1), and $\omega$ (from 0--2$\pi$), and uniform improper priors on all other parameters. In this way, we are sampling only values of $a/R_{\star}$ that are consistent with the spectroscopically-derived parameters, but values of $a/R_{\star}$ that are not as well matched to the light curve are penalized by the $\chi^{2}$ term corresponding to the photometry. We adopt Gaussian priors on  $R_{\star}$ and $T_{\mbox{eff}}$, which we implement by adding extra terms in the $\chi^{2}$ (where {\bf P} corresponds to the vector of light curve parameters at each iteration): 

\begin{equation}
\chi^{2}=\sum_{i=1}^n\left(\frac{f_{i}-m({\bf P})_{i}}{\sigma_{i}}\right)^{2}+ \left(\frac{\Delta T_{\rm eff}}{\sigma_{T_{\rm eff}}}\right)^{2}+\left(\frac{\Delta R_{\star}}{\sigma_{R_{\star}}}\right)^{2}.
\label{eq:chisquared_5}
\end{equation}

\begin{figure}
\begin{center}
 \includegraphics[width=8in,angle=90]{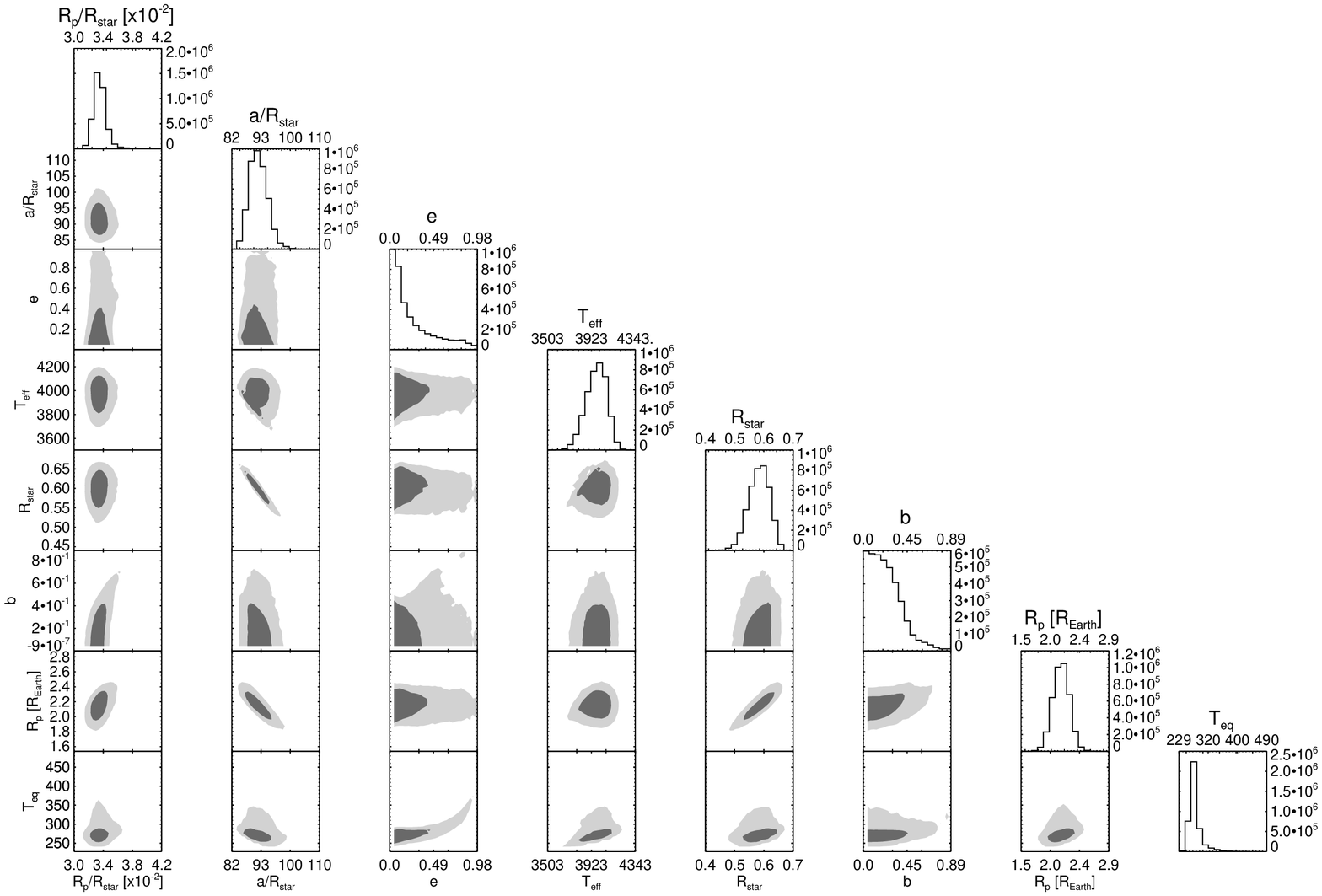} 
 \caption{Markov Chain Monte Carlo probability distributions for light curve parameters of Kepler-61. The dark gray area encloses 68\% of the values in the chain, while the light grey area encloses 95\% of the values. We assign the range of values corresponding to 1$\sigma$ confidence from the area enclosing 68\% of the values nearest to the mode of the posterior distribution for each parameter (as described in the text).}
\end{center}
\label{fig:mcmc_results_5}
\end{figure}

\begin{figure}[h!]
\begin{center}
 \includegraphics[width=5in]{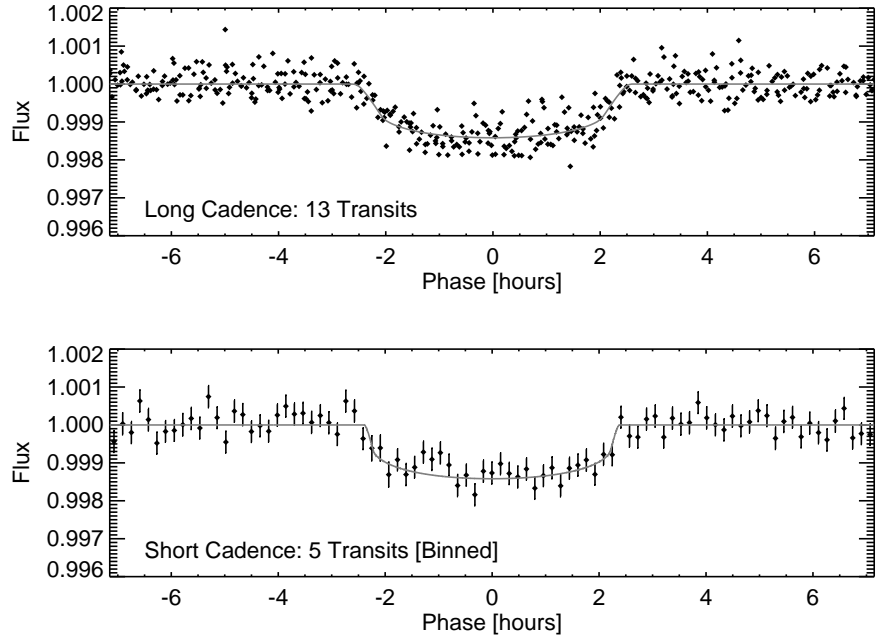} 
 \caption{{\it Top panel:} Kepler-61 transit light curve as a function of phase for Quarters 1--11, gathered in long-cadence observing mode. {\it Bottom panel:} \kepler\ light curve for Quarters 12, 13, and 14, gathered in short-cadence observing mode and binned in intervals of 7 minutes. Overplotted in gray is the best transit model light curve, with parameters given in Table 2. The effect of integrating over the 29.5 minute exposure time corresponding to long-cadence observations is apparent in the different shapes of transit during ingress and egress.}
  \label{fig:keplerfit5}
\end{center}
\end{figure}

In Figure \ref{fig:mcmc_results_5}, we show the correlations between the posterior distributions of subset of parameters in the model fit, as well as the histograms corresponding to each parameter. In Figure \ref{fig:keplerfit5}, we show the phased \kepler\ transit light curve for Kepler-61b, with the best-fit transit light curve overplotted. We report the best-fit parameters and uncertainties in Table 2. The range of acceptable solutions for each of the light curve parameters is determined as follows. Following \cite{Torres08}, we report the most likely value from the mode of the posterior distribution, marginalizing over all other parameters. The uncertainty is derived from the extent of the posterior distribution that encloses 68\% of values closest to the mode.  

\begin{figure}
\begin{center}
 \includegraphics[width=5in]{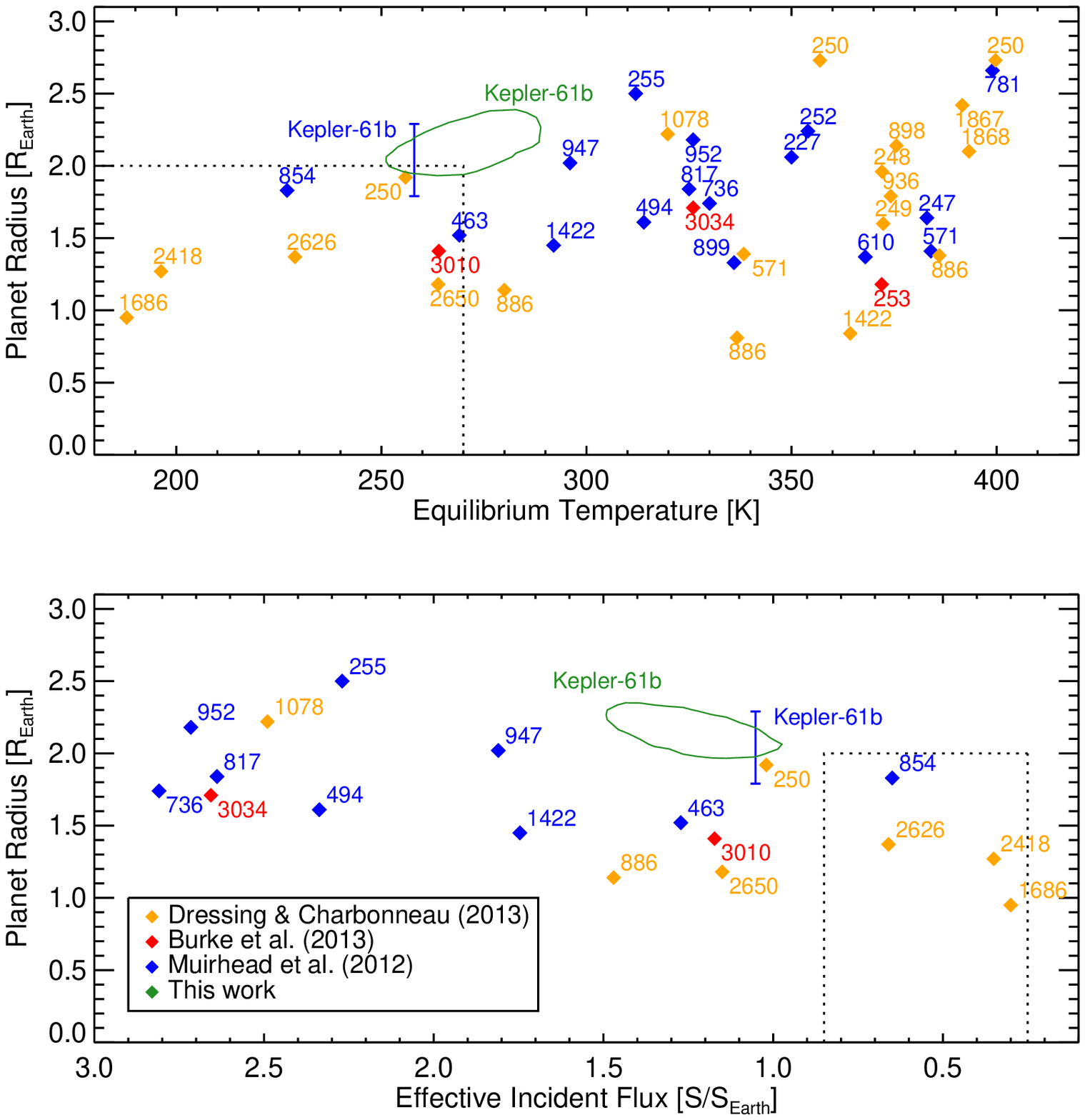} 
 \caption{({\it Top panel}): Sample of \kepler\ candidate exoplanets with radii $<3$ $R_{\oplus}$, orbiting stars with $T_{\mbox{eff}}<$4200 K, and with equilibrium temperatures $<400$ K.  The dotted lines denote 270 K  (above which runaway greenhouse effect occurs, per \citealt{Kaltenegger11}), and 2 $R_\oplus$. ({\it Bottom panel}) Sample of \kepler\ candidate exoplanets with radii $<3$ $R_{\oplus}$, orbiting stars with $T_{\mbox{eff}}<$4200 K, and incident flux levels within 3 times the value received at the surface of the Earth. The dotted lines here denote the habitable zone for stars from 2700-4500 K, per \cite{Kopparapu13}. The number labels adjacent to points depict the KOI number. The values for planet candidates calculated by \cite{Muirhead12b} orbiting stars with T$_{\mbox{eff}}<$3800 K are depicted in blue. For candidates orbiting hotter stars, for which the $K$--band method tends to underpredict temperature, we depict  values instead from \cite{Burke13} in red (this stellar characterization relies instead upon the comparison of the broadband colors to stellar models of \citealt{Pinsonneault12}). Where these latter values have been revised by \cite{Dressing13} using the Dartmouth stellar models \citep{Dotter08}, we have used those radius and temperature values and depicted the KOI in gold. The radius and temperature for Kepler-61 (KOI 1361) from $K$-band spectroscopy reported by \cite{Muirhead12b} is given by the blue error bar, while the revised 1$\sigma$ contour from this work is shown in green.}
  \label{fig:isochrone_aonr}
\end{center}
\end{figure}

\subsection{Physical Parameters}
This procedure described in Section 1.1 is also advantageous in that, in addition to recording the value of $a/R_{\star}$ at each iteration of the MCMC chain, we may also record the other traits of the star at that value of $R_{\star}$ and $T_{\mbox{eff}}$, including its mass, luminosity, and age. At the conclusion of the MCMC analysis, therefore, we have accumulated a chain not only for the light curve parameters, but for the physical parameters as well, as predicted from evolutionary models. The correlations between parameters, both physical parameters and those associated with the light curve, are therefore preserved in the chain and incorporated into our estimate of the stellar parameters (although we note that we have not accounted for possible correlated error between our adopted values of effective temperature and radius measurements of the star). We calculate the planetary radius from multiplying the elements of the $R_{p}/R_{\star}$ by the chain of $R_{\star}$. We infer a value for the stellar mass from its posterior distribution of $M_{\star}$=0.635$\pm$0.037. The slowly-evolving nature of Kepler-61 results in a largely unconstrained estimate of stellar age. The stellar rotation period (36 days, described in Section 6.1) indicates a star older than 1 Gyr, which is consistent, though also itself only a weak constraint. It's also possible to evaluate the posterior distribution of planetary equilibrium temperatures from the MCMC analysis. In the case of a circular orbit, we require only the stellar radius and temperature, the planetary semimajor axis, and the planetary albedo. However, in the case of an eccentric orbit, the planet receives time-variable stellar insolation. In order to evaluate the equilibrium temperature of the planet in the case of non-zero eccentricity, we evaluate the time-averaged equilibrium temperature by performing an integral over the mean anomaly from 0 to 2$\pi$, using the formalism detailed in \cite{Murray10}, where $A$ is the geometric albedo of the planet, $d$ is its distance from the star, and $M$ is the mean anomaly:

\begin{equation}
\frac{1}{2\pi}\int_{0}^{2\pi} T_{eq}\mbox{d}M=\frac{1}{2\pi}(1-A)^{1/4}\int_{0}^{2\pi} \sqrt{\frac{R_{\star}}{2d}}T_{\star}\mbox{d}M
\label{timeaverage}
\end{equation}

\begin{figure}[h!]
\begin{center}
 \includegraphics[width=5in]{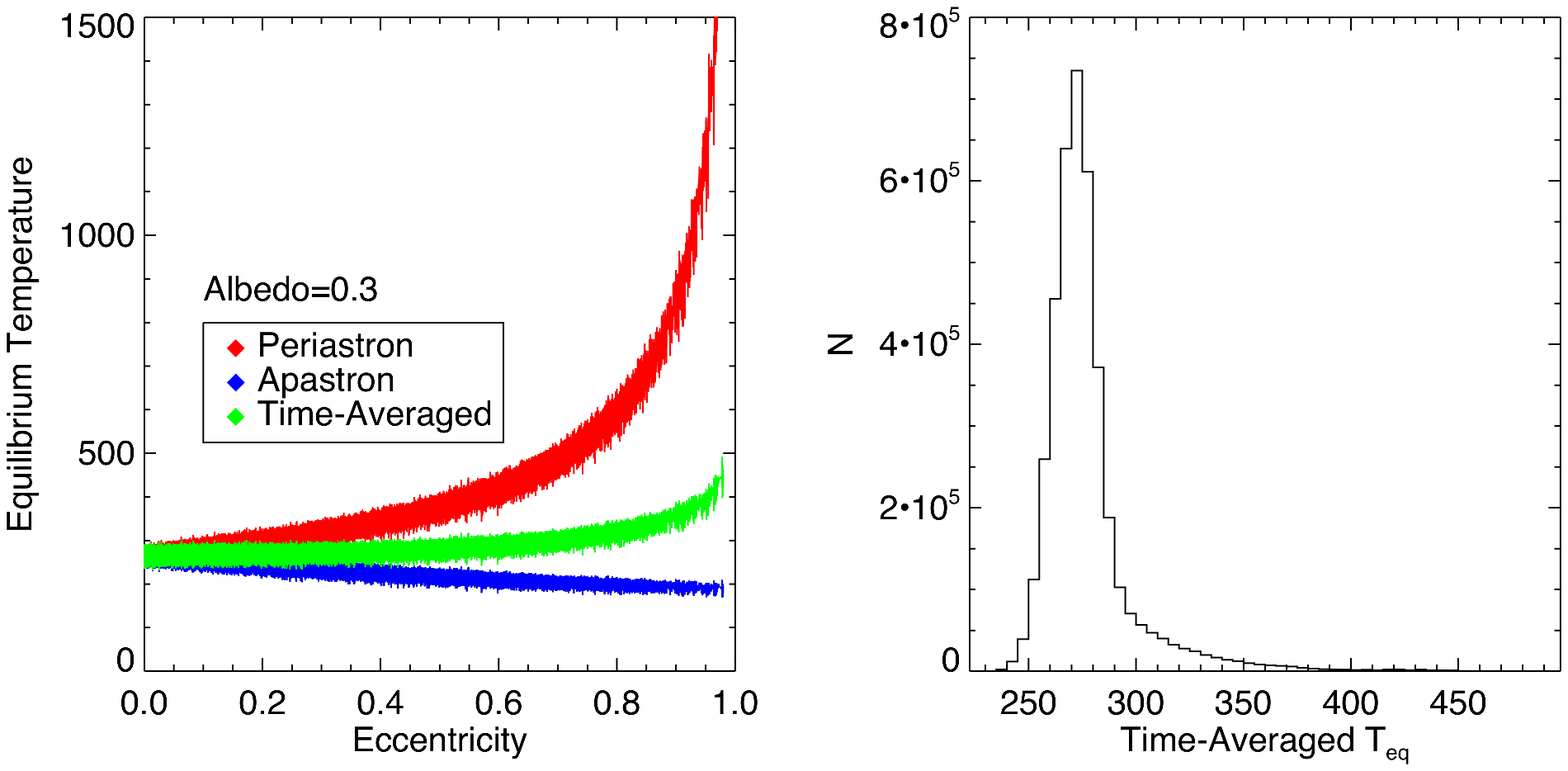} 
 \caption{{\it Left:} For each iteration of the MCMC chain, we calculate the temperature of the planet at apastron (blue), periastron (red) and the time-averaged equilibrium temperature (green). We have assumed an albedo of 0.3. {\it Right:} Histogram of time-averaged equilibrium temperatures for Kepler-61b. The long tail of temperatures higher than 350 K is contributed entirely from orbital eccentricities $>0.8$.}
  \label{fig:temp}
\end{center}
\end{figure}

In Figure \ref{fig:temp}, we show both the planetary temperatures based on the apastron and periastron isolation from the star, as well as the time-averaged temperature of the planet, for each element of the MCMC chain.

We imposed a flat prior on the eccentricity of Kepler-61 from 0--1, so the posterior distribution on $e$ includes values as high as 0.9 (such a large eccentricity still matches the transit duration at finely-tuned values of $\omega$). Indeed, the circularization timescale for Kepler-61 has not yet elapsed if we assume it did not possess a large initial eccentricity. We consider the expression for circularization time (for modest initial $e$) given by \cite{Goldreich66}, where $a$ is the semimajor axis of the planet, $R_{p}$ is the planetary radius, $M_{p}$ is the planetary mass, $M_{\star}$ the stellar mass, $Q$ is the tidal quality factor for the planet (which is highly uncertain, but we test two values: 100 for the assumption of a terrestrial composition, per \citealt{Goldreich66}, or $10^{4}$, which is the lower limit measured by \citealt{Banfield92} for Neptune) and $G$ is the gravitational constant:

\begin{equation}
t_{circ}=\frac{4}{63}\frac{1}{\sqrt{GM_{\star}^{3}}}\frac{M_{p}a^{13/2}Q}{R_{p}^{5}}
\end{equation}

We find that the circularization timescale would be 400 Myr for a terre strial $Q$ of 100 (assuming a 7 $M_{\oplus}$ planet orbiting a 0.64 $M_{\odot}$ star at 0.25 AU), and 190 Gyr for a Neptune-like $Q$ of 10$^{4}$. It's therefore plausible that the planet resides in an eccentric orbit. If Kepler-61 began with a large initial eccentricity, then terms of order (1-$e^{2}$) become relevant and the circularization timescale decreases (as elucidated in Equations 7--9 of \citealt{Socrates12}). It's therefore also possible that the circularization timescale has indeed elapsed, dependent upon the initial eccentricity and tidal $Q$ of the planet. However, the uncertainty of our knowledge about its initial eccentricity (coupled with the uncertainty about the correct value of tidal $Q$ for Kepler-61 in particular) is such that we believed a flat prior on $e$ to be appropriate. 

\begin{deluxetable*}{rr}
\tabletypesize{\scriptsize}
\singlespace
\tablecaption{Star and Planet Parameters for Kepler-61}
\label{tbl:params_5}
\tablewidth{0pt}
\tablehead{
\colhead{Parameter} & \colhead{Value \& 1$\sigma$ confidence interval} \\}
\startdata
Kepler-61 [star]& \\
\hline
Right ascension\tablenotemark{a} & 19h41m13.09s \\
Declination\tablenotemark{a} & +42d28m31.0s \\ 
$T_{\mbox{eff}}$ [K]\tablenotemark{b}&        4017$^{+68}_{-150}$  \\
$R_{\star}$ [Solar radii]\tablenotemark{b} &  0.62$^{+0.02}_{-0.05}$ \\
$M_{\star}$ [Solar masses] &    0.635$\pm$0.037 \\
$\mbox{[Fe/H]}$\tablenotemark{c} &   0.03$\pm$0.14 \\
Age [Gyr] & $>$1  \\
\hline
Kepler-61 [planet]& \\
\hline
Period [Days] &   59.87756$\pm$0.00020 \\
$T_{0}$ [BJD-2450000] &        4984.1880$^{+0.0029}_{-0.0024}$ \\
$R_{p}/R_{\star}$ &      0.03301$\pm$0.00085 \\
a/$R_{\star}$ &        90.6$\pm$3.4 \\
inc [deg] &        $>$89.80 \\
e &    $<$0.25 \\
e cos($\omega$) &     0.0$\pm$0.29 \\
Impact Parameter &   $<$ 0.29 \\
Total Duration [min] &        290.7$\pm$4.6 \\
Ingress Duration [min] &        9.56$\pm$0.47 \\ 
$R_{p}$ [Earth radii] &        2.15$\pm$0.13 \\
Planetary $T_{eq}$ [K] &        273$\pm$13 \\
\enddata
\tablenotetext{a}{ICRS (J2000) coordinates from the 2MASS All-Sky Catalog of Point Sources \citep{Cutri03}. The proper motion derived by \cite{Roeser10} is -5.6 milliarcsec yr$^{-1}$ in right ascension and 11.8 milliarcsec yr$^{-1}$ in declination (both with error bar of 3.8 milliarcsec yr$^{-1}$).}
\tablenotetext{b}{Stellar temperature and radius inferred from weighted mean of directly measured K7 and M0 sample, as described in text.}
\tablenotetext{c}{Metallicity derived from $K$--band using metric from \cite{Mann12}.}
\end{deluxetable*}

We find consistent orbital parameters for the 11 quarters of long-cadence observations (13 transits) and the three quarters of short cadence observations (5 transits), though our ability to resolve the shape of ingress and egress with short-cadence mode, coupled with the prior on $a/R_{\star}$ from our knowledge of the stellar parameters, is reflected in the error bars on $a/R_{\star}$ and the ingress/egress time $\tau$, even though we have only half the number of transits in the latter mode. While we find $\tau$=9.3$\pm$3.2 minutes from the 13 transits in long-cadence mode, we find $\tau$=9.56$\pm$0.47 minutes from the five transits in short-cadence mode.

\section{Planetary Validation of Kepler-61}

\cite{Morton11} provide {\it a priori} false positive probabilities for the \kepler\ planetary candidates published by \cite{Borucki11}, within which sample Kepler-61 is included. They cite the vetting of candidates by the \kepler\ software (detailed by \citealt{Batalha10}) as being already sufficient to produce a robust list of candidates, and combine stellar population synthesis and galactic structure models to demonstrate that nearly all of these 1235 candidates have a false positive probability $<$10\%. Kepler-61b, with a \kepler\ magnitude of 14.995 and a galactic latitude of 9.6$^{\circ}$, has an  {\it a priori} false positive probability of 4.8\%. 

\subsection{Adaptive Optics Imaging}
\label{sec:ao_5}

\begin{figure}[h!]
\begin{center}
 \includegraphics[width=5in]{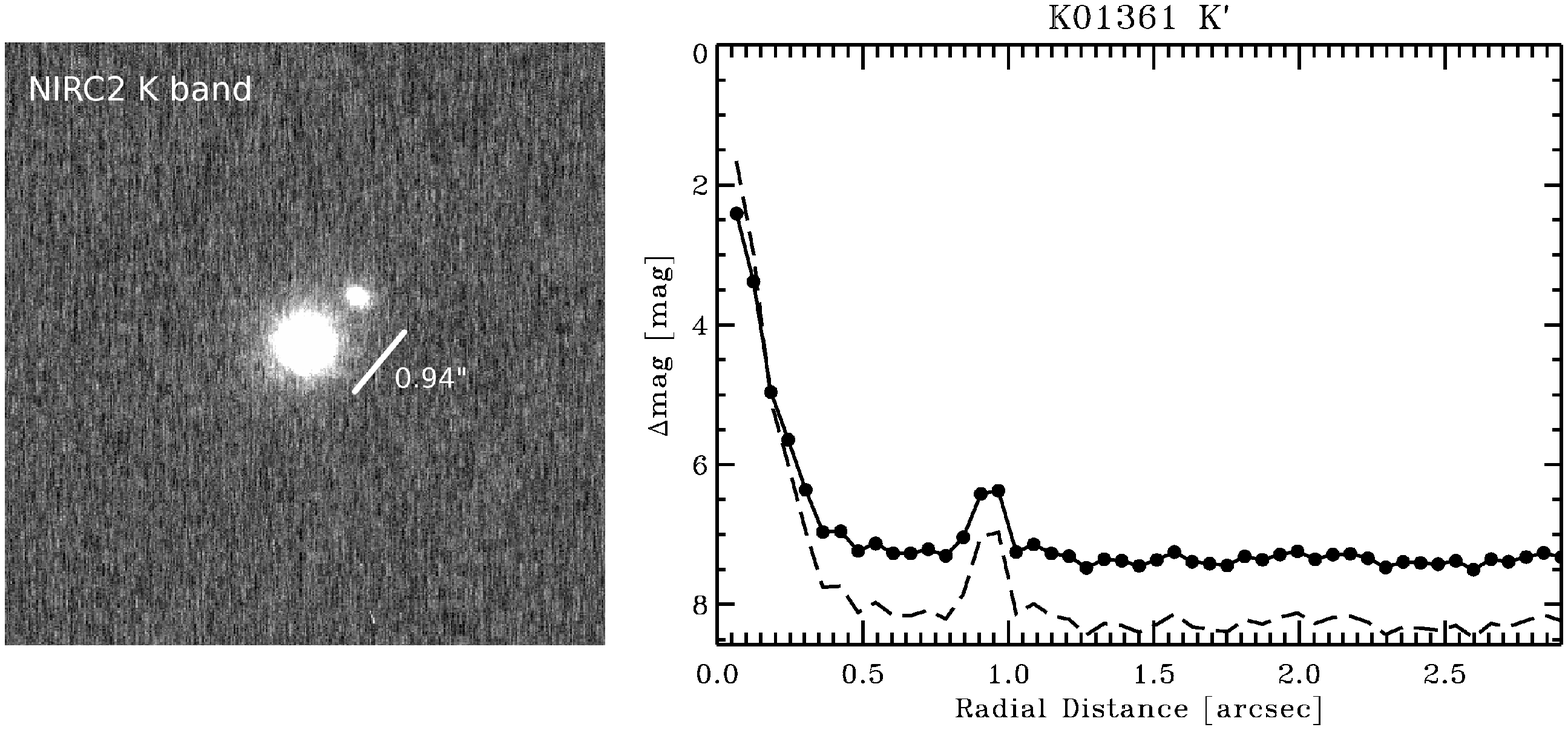} 
 \caption{{\it Left:} $K$ band adaptive optics image of Kepler-61. An additional companion is located 0.94'' away from the target star, and is 2.9 magnitudes fainter. {\it Right:} The sensitivity limits to additional point sources in the neighborhood of Kepler-61 as a function of radial distance from the primary target.  The filled circles represent the $K$ magnitude limits and each point represents a step in FWHM away from the primary target centroid peak.  The dashed line underneath represents the $K$-band limits converted to \kepler\ magnitude limits if a star were to have a nominal $Kp$-$K$ color, as described in the text.}
  \label{fig:AO}
\end{center}
\end{figure}

We place limits on the presence of additional stars in the neighborhood of Kepler-61 with adaptive optics (AO) observations gathered at Keck with the NIRC2 instrument. On 22 June 2012 we observed Kepler-61 in both $J$ and $K$ band, with observed FWHM of the core in $K$ of 0.06'' and in $J$ of 0.10". We detect a companion star, 2.9 magnitudes fainter and located 0.94'' to the northwest of Kepler-61, which is shown in Figure \ref{fig:AO}.  The high resolution $K$-band AO image cleanly resolves these two sources, while we employed the $J$-band image to better characterize the neighboring source. The additional source falls within the \kepler\ aperture, but is removed enough from the target star that we can employ the \kepler\ centroids to assert that the planet orbits the brighter star, as we describe in Section \ref{sec:centroid_5}. The blended magnitudes are 15.0 in the \kepler\ bandpass, $J$=13.077$\pm$0.022 magnitudes and $K$= 12.272$\pm$0.019 magnitudes. We independently measure the $J$ and $K$ magnitudes for the two stars, and apply the Kp-$K$ and $J$-$K$ relationships (which are derived separately for dwarfs and giants from the \kepler\ Input Catalog; this conversion is described in detail in Appendix A of \citealt{Howell12}) to determine the de-blended \kepler\ magnitude. For the primary target, we find Kp = 15.22$\pm$0.09 mag, $J$ = 13.149$\pm$0.022 magnitudes, and  $K$ = 12.345$\pm$0.019 mag. For the secondary, we find Kp= 18.20$\pm$0.10 mag, $J$=16.064$\pm$0.025 mag, and $K$=15.242$\pm$0.020 mag. The two stars exhibit indistinguishable $J-K$ colors, with 0.804 $\pm$ 0.029 for Kepler-61 and 0.822$\pm$0.32 for the dimmer companion. 

We assess our sensitivity to additional sources using a similar procedure to that described by \cite{Batalha11}. We inject fake sources near the target star at random position angles, using steps in magnitude of 0.5 mag and varying the distance from the target star in increments of 1.0 FWHM of the point-spread function (PSF). We then attempt to identify the injected sources with the DAOPhot routine \citep{Stetson87} and also by eye, and set our sensitivity limit, as a function of distance, at the magnitude where we are able to recover the injected sources. The limit in $\Delta m$ as a function of distance from the target star is shown in Figure \ref{fig:AO}. We then convert the $\Delta m$ sensitivity limit in $K$ band to a limit in \kepler\ magnitudes, again using the $Kp$-$K$ relationship detailed in \cite{Howell12}.

\subsection{Speckle Imaging}
We gathered speckle imaging of Kepler-61 on 11 June 2011 UT using the two-color Differential Speckle Survey Instrument at the Wisconsin Indiana Yale NOAO (WIYN) 3.5 m telescope, located at Kitt Peak Observatory \citep{Horch09}. The speckle camera obtained 7000 40 msec images in $I$ band (8880/400\AA). We reduced and processed these observations to produce a final reconstructed speckle image for the star. Details of the speckle camera observations for the \kepler\ follow-up observing program, including the reduction methods, are presented in \cite{Howell11}.

The speckle observations allow detection of a companion star within the approximately $2.76 \times 2.76$ arcsec box centered on the target.  We can detect, or rule out, companions between 0.05 arcsec and 1.8 arcsec from Kepler-61 and, in this case, we did not detect a companion star. We report the limiting difference in magnitude for an additional star that would have been detectable with 3$\sigma$ confidence in Table 3. The companion detected with adaptive optics imaging, which we describe in the previous section, lies just beyond detectability with speckle imaging in $I$ band, at a distance of 0.94'' and $\Delta m$ in $K$ band=2.9 (0.25 magnitudes from what would have been detected in the speckle image with 3$\sigma$ confidence at that distance from the star).

\begin{deluxetable*}{cc}
\tabletypesize{\scriptsize}
\singlespace
\tablecaption{Magnitude Limits on Companions to Kepler-61 from Speckle Imaging}
\label{tbl:speckle}
\tablewidth{0pt}
\tablehead{
\colhead{Radius of Annulus around Kepler-61} & \colhead{Limiting Delta Magnitude} \\
\colhead{[Arcseconds]} & \colhead{3$\sigma$ Confidence} \\}
\startdata
   0.05--0.30 & 2.69 \\
   0.30--0.50 & 3.05 \\
   0.50--0.70 & 3.09 \\
   0.70--0.90 & 3.16 \\
   0.90--1.10 & 3.15 \\
   1.10--1.30 & 3.11 \\
   1.30--1.50 & 3.18 \\
   1.50--1.70 & 3.24 \\
   1.70--1.90 & 3.20 \\
\enddata
\end{deluxetable*}

\subsection{Photocenter Tests}
\label{sec:centroid_5}

We use two methods to search for false positives due to background eclipsing binaries, based on examination of the pixels in the aperture of Kepler-61: direct measurement of the source location via difference images, and inference of the source location from photocenter motion associated with the transits.  We employ two methods because of their different vulnerabilities to systematic bias; when the methods agree, we have increased confidence in their result.

Difference image analysis \citep{Torres11a} takes the difference between average in-transit pixel images and average out-of-transit images.  A fit of the Kepler pixel response function (PRF; \citealt{Bryson10})  to both the difference and out-of-transit images directly provides the location of the transit signal relative to the host star.  We measure difference images separately in each quarter, and estimate the transit source location as the robust uncertainty-weighted average of the quarterly results.

We measure photocenter motion by computing the flux-weighted centroid of the pixels in the optimal aperture, plus a 
one-pixel halo in every cadence, generating a centroid time series for row and column. We fit the modeled transit to the whitened centroid time series transformed into sky coordinates.  We perform a single fit for all quarters, and then infer the source location by scaling the difference of these two centroids by the inverse of the flux as described in \cite{Jenkins10b}.  

The source as determined by the difference image method is offset from the nominal location of Kepler-61, as given in the \kepler\ Input Catalog, by 0.09 $\pm$ 0.29 arcsec $= 0.68 \sigma$. The source as determined by the flux-weighted centroid method is offset from Kepler-61 by 0.32 $\pm$ 0.37 arcsec $= 0.86 \sigma$.  Both methods show that the observed centroid location is consistent with the transit occurring at the location of Kepler-61, and rule out the companion in the adaptive optics imaging as the source of the transit, which is $3\sigma$ removed from position at which the transit occurs.

\subsection{{\it Spitzer} Observations}
Warm \spitzer\ observations in the near-infrared can also prove useful toward validating \kepler\ candidates, as shown for Kepler-10c \citep{Fressin11}, Kepler-14b \citep{Buchave11}, Kepler-18c \& d \citep{Cochran11}, Kepler-19b \citep{Ballard11b}, Kepler-22b \citep{Borucki11}, Kepler-25b \& c \citep{Steffen12}, and Kepler-20c, d, e, \& f \citep{Gautier12, Fressin12} Unless a putative blend scenario is comprised of stars of nearly identical color, the transit depth in a blend scenario will depend upon the wavelength at which it is observed. Conversely, an authentic transiting planet will produce an near-achromatic transit depth.

We gathered observations using the Infrared Array Camera (IRAC) \citep{Fazio04} on Warm {\it Spitzer} at 4.5~$\mu$m of the UT 17 September 2011 transit of Kepler-61b. The observations span 10 hours, centered on the 4.75-hour-long transit. We gathered the observations using the full-array mode of IRAC, with an integration time of 12 s/image. We employed the techniques described in \cite{Agol10} for the treatment of the images before photometry. We first converted the Basic Calibrated Data products from the {\it Spitzer} IRAC pipeline (which applies corrections for dark current, flat field variations, and detector non-linearity) from mega-Janskys per steradian to data number per second, using 0.1469 MJy$\cdot$sr$^{-1}$ per DN s$^{-1}$, and then to electrons per second, using the gain of 3.71 $e$ DN$^{-1}$. We identified cosmic rays by performing a pixel-by-pixel median filter, using a window of 10 frames. We replace pixels that are $>4\sigma$ outliers within this window with the running median value. We also corrected for a striping artifact in the Warm {\it Spitzer} images, which occurred in the same set of columns, by taking the median of the pixel values in the affected columns (using only rows without an overlying star) and normalizing this value to the median value of neighboring columns. 
 
We discuss several means of performing the Warm \spitzer\ IRAC photometric reduction for similar observations in \cite{Ballard11b}, and make use of the conclusions from that work. First, we estimate the position of the star on the array with a flux-weighted sum of the position within a circular aperture of 3 pixels. We then performed aperture photometry on the images using the centroid positions and variable aperture sizes between 2.1 and 4.0 pixels, in increments of 0.1 pixels up to 2.7 pixels, and then at 3.0 and 4.0 pixels. We decided to use the position estimates using a flux-weighted sum at an aperture of 2.6 pixels, which minimized the out-of-transit RMS. 

We remove the effect of the IRAC intrapixel sensitivity variations, or the ``pixel-phase'' effect (see eg. \citealt{Charbonneau05, Knutson08}) using a polynomial functional form for the intrapixel sensitivity (which depends upon the $x$ and $y$ position of the star on the array). We denote the transit light curve $f$ (which depends upon time), and we hold all parameters constant except for the transit depth. We use the light curve software of \cite{Mandel02} to generate the transit models. The model for the measured brightness $f'(x,y)$ is given by:

\begin{equation}
f'=f(t,R_{p}/R_{\star})\cdot[b_{1}+b_{2}(x-\bar{x})+b_{3}(x-\bar{x})^{2}+b_{4}(y-\bar{y})+b_{5}(y-\bar{y})^{2}],
\label{eq:spitzerpoly_5}
\end{equation}

where we include all of the observations (both in- and out-of-transit) to fit the polynomial coefficients and the transit depth simultaneously.

We fit for the polynomial coefficients $b_{1}$ through $b_{5}$ using a Levenberg-Marquardt $\chi^{2}$ minimization. However, the Spitzer light curve contains significant correlated noise even after the best intrapixel sensitivity model is removed. We incorporate the effect of remaining correlated noise with a residual permutation analysis of the errors as described by \cite{Winn08}, wherein we find the best-fit model $f'$ to the light curve as given by Equation \ref{eq:spitzerpoly_5}, subtract this model from the light curve, shift the residuals by one data point in time, add the same model back to the residuals, and refit the depth and pixel sensitivity coefficients. We wrap residuals from the end of the light curve to the beginning, and in this way we cycle through every permutation of the data. We determine the best value from the median of this distribution, and estimate the error from the closest 68\% of values to the median. Using the residual permutation method on the light curve treated with a polynomial, we find $R_{p}/R_{\star}$=0.0315$\pm$0.0069, in excellent agreement with the \kepler\ measurement of $R_{p}/R_{star}$=0.03476$\pm$0.00094.

\begin{figure}[h!]
\begin{center}
 \includegraphics[width=5in]{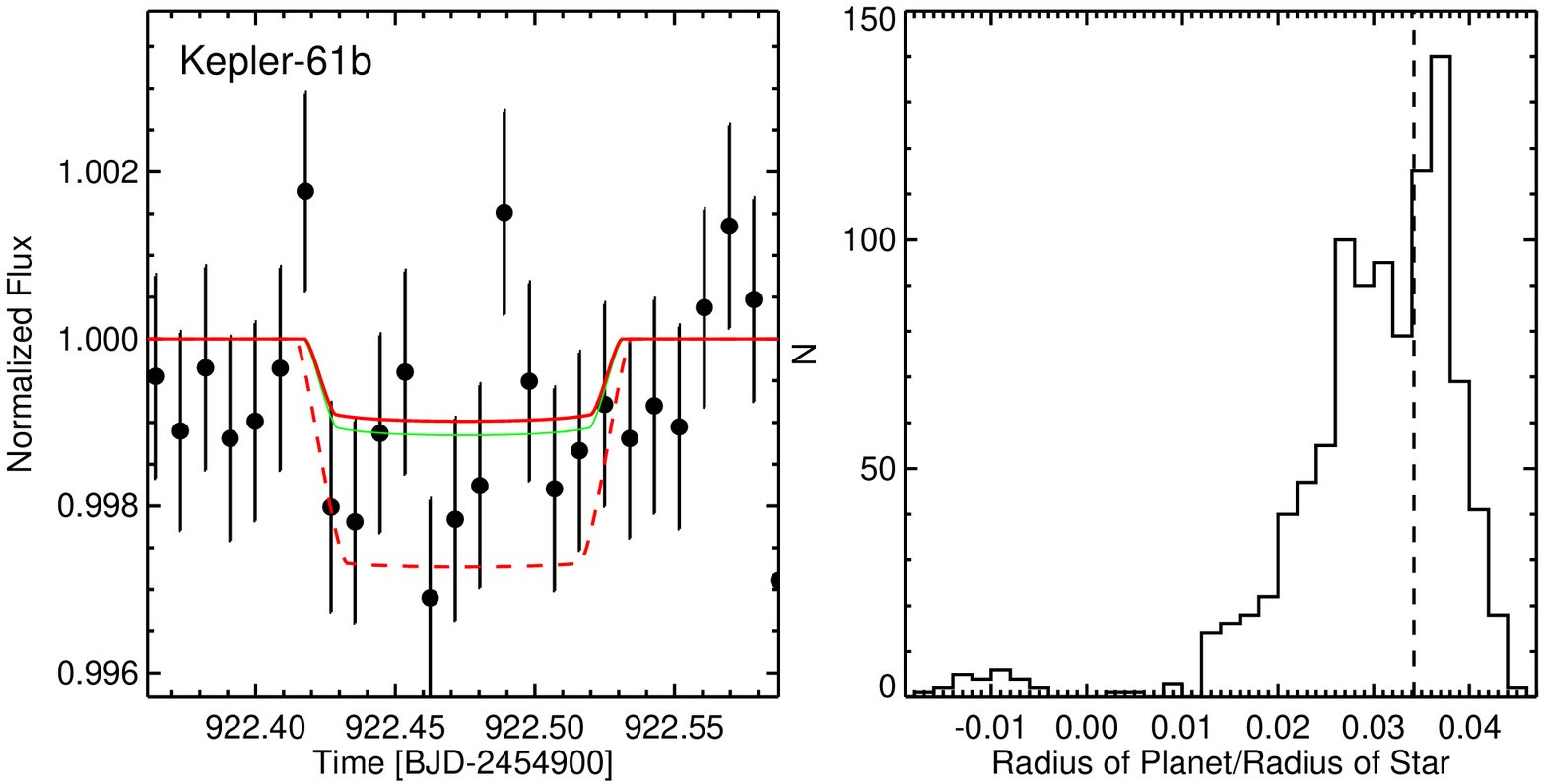} 
 \caption{{\it Left:} Transit of Kepler-61b gathered with Warm \spitzer\ at 4.5 $\mu$m, binned by a factor of 16. The best-fit transit model with depth derived from the \spitzer\ observations is shown with a solid red line, while the \kepler\ transit model (with \spitzer\ 4.5 $\mu$m channel limb darkening) is shown in green. The \spitzer\ and \kepler\ transit depths are in excellent agreement. The transit depth we can rule out with 3$\sigma$ confidence at 4.5 $\mu$m is shown by a dashed red line. {\it Right:} The results of a residual permutation analysis on the \spitzer\ transit of Kepler-61b. We detect the transit with 3$\sigma$ confidence, and the depth inferred from the \kepler\ light curve, indicated by a dotted line, lies within one standard deviation of the depth inferred from \spitzer.}
  \label{fig:spitzer_5}
\end{center}
\end{figure}

We note that the use of the weighted sensitivity function proposed in \cite{Ballard10b} made a negligible difference to the photometric residuals in this case, so for reasons of computational time, we deferred to the polynomial reduction technique. In Figure \ref{fig:spitzer_5}, we show the combined and binned \spitzer\ light curve, with the best-fit transit model derived from the \spitzer\ observations and the best-fit \kepler\ transit model (with the quadratic limb darkening coefficients for the \spitzer\ 4.5 $\mu$m filter, drawn from \citealt{Claret11} as similarly described in Section 1) overplotted. 


\subsection{\blender\ Validation}

Traditional confirmation of the planetary nature of a transit signal relied upon a dynamical mass measurement from radial velocity observations. In cases where the reflex motion induced on the host star by the planet is too small to be detected, dynamical confirmation may yet be possible via transit timing variations (TTVs). However, in cases where a dynamical mass measurement via either technique is not viable, it's still possible to ``validate'' the planetary nature of the transit signal, via a statistical argument about the relative likelihood of an authentic planet producing the transit signal, as compared to a false positive scenario. \blender\ is such a machinery, which combines evidence from the \kepler\ photometry (as compared to model light curves for planetary and false positive scenarios), spectroscopy, \spitzer\ photometry (where available), the stellar colors, and adaptive optics to deduce a false positive probability for a planetary candidate. \blender\ has already been applied to validate planets in a number of \kepler\ exoplanetary systems, including CoRoT-7 \citep{Fressin12}, Kepler-10 \cite{Fressin11}, Kepler-18 \citep{Cochran11}, Kepler-19 \citep{Ballard11b}, Kepler-20 \citep{Gautier12}, Kepler-21 \citep{Howell12}, and Kepler-22 \citep{Borucki12}, and its details are described therein as well as in \cite{Torres04}, \cite{Fressin11}, and \cite{Fressin12}. We summarize the \blender\ procedure below, and present the statistical likelihood that the transit signal presented by Kepler-61b is attributable to a 2.5 $R_{\oplus}$ planet orbiting a 0.65 $R_{\odot}$ star.

The exquisite precision of the \kepler\ photometry is already sufficient to rule out some false-positive scenarios, which would produce a significantly different transit shape from the one observed by \kepler. Such a false positive could mimic the observed transit depth if an additional star fell within the same aperture of the \kepler\ target star. The light contributed by this undetected companion (which may be gravitationally bound to the target star or lie in the foreground or background), would reduce the transit depth produced by an eclipsing binary system or a planetary system comprising a star and a larger planet, conspiring to produce a planetary transit depth. \blender\ manufactures synthetic light curves corresponding to these false positive scenarios: for those with a physically bound companion, \blender\ assumes a common age for the putative companion star and the \kepler\ target star, while an unassociated background or foreground star is assigned an age of 3 Gyr. The mass of this secondary star and the tertiary body (either star or planet) is allowed to vary. These model blend light curves are compared to the \kepler\ photometry in a $\chi^{2}$ sense. Blend scenarios that furnish a good fit to the \kepler\ light curve (within 3$\sigma$ of the best authentic planet model) are then tested for consistency against other constraints. These include (a) the color of the star as reported in the KIC \citep{Brown11}, which allows us to rule out any simulated blends resulting in a combined color that is significantly redder or bluer than the target; (b) limits from the centroid motion analysis on the angular separation of
companions that could produce the signal (Section \ref{sec:centroid_5}); (c) brightness and angular separation limits from high-resolution adaptive optics (Section \ref{sec:ao_5}); and (d)
constraints from the measured transit depth derived from our \spitzer\ observations, which place an upper limit on the mass (spectral type) of stars producing the blend. For the hierarchical triple scenario (in which the secondary star is physically bound to the \kepler\ target star), we considered dynamical stability constraints \citep{Holman99}.  Surviving blend scenarios that satisfy all of the above criteria are folded into the ``blend frequency'': which is the probability that such a finely-tuned blend lies near enough to the target star to be undetectable by adaptive optics imaging (using the sensitivity limits we find in Section \ref{sec:ao_5}). We compared this frequency with the expected frequency of true planets (the planet ``prior'') to derive the ``odds ratio''. To estimate the planet prior, we employ the list of candidate planets (KOIs) from \cite{Batalha13}, restricted to main-sequence host stars. We assume that this list is complete (i.e., that all signals have been detected) and that the rate of false positives is negligible (which assumption is justified by the findings of \citealt{Morton11}, who found a false-positive rate of $<5$\% for most KOIs).

We find that background eclipsing binaries comprising two stars furnish only poor fits to the \kepler\ photometry, and so are excluded from \kepler\ data alone. A portion of the hierarchical triple parameter space (with a star and larger planet gravitationally bound to the \kepler\ target star) provides good fits to the \kepler\ photometry, but these scenarios are then excluded by the combination of observational constraints described above. Background stars transited by larger planets, on the other hand, can mimic the \kepler\ photometry and remain consistent with the \kepler\ centroid motion, follow-up adaptive optics imaging, spectroscopy, and \spitzer\ constraint. We find that the frequency of background/foreground blends that satisfy these criteria is 4.19$\times10^{-8}$. The planet prior is estimated by counting the number of known KOIs that are in the same (3$\sigma$) radius range as the putative planet (105 in this
case), and dividing by the total number of main-sequence \kepler\ targets observed during Q1-Q6 (138,253). We obtained a planet prior of 105 / 138,253 = 7.60$\times10^{-4}$.  We conclude that a true transiting planet is 7.60$\times10^{-4}$ / 4.19$\times10^{-8}$ = 18,000 times more likely than a blend, which allows us to validate Kepler-61 with a  high degree of confidence.

The simple procedure described above for estimating the planet prior does not take into account the period of the signal, which may be an important factor for small and long-period candidates (such as Kepler-61b) because such signals are rare (see below). Furthermore, the completeness and purity of the KOI list of
Batalha et al. (2012), on which our planet prior calculation relies, may decline with period and planetary radius, whereas we have assumed these concerns are negligible. These factors may in principle influence both the planet prior and the blend frequencies we have just described (since we employ the KOI list not only to estimate the occurrence of authentic small planets, but also to estimate the rate of occurrence of larger planets in false-positive blend scenarios). Therefore, instead of allowing eclipsing binaries and transiting planets with any orbital period to factor into the blend frequency calculation, we elected to redo the \blender\ analysis with a more realistic approach to allowed blends. First, we accept only blends with periods near the measured periods of Kepler-61 (within a factor of two) for both the blend and planet prior calculation. To address the concerns about completeness and purity of the KOI list, we performed separate Monte Carlo simulations to establish incompleteness corrections for the KOI list and also to estimate the false positive rates for planets in the size ranges relevant to this calculation. A description of this work is forthcoming (Fressin et al., in preparation). We obtained a revised frequency of background/foreground blends of 7.27$\times10^{-9}$.  Examining the candidate list of \cite{Batalha13}, we found 22 KOIs in the relevant radius range with periods within a factor of two of the period of Kepler-61. Our simulations suggest that about 2.09 of these may be false positives, but also that the KOI list for signals of this size and period is in fact incomplete, requiring a correction factor of approximately 1.59 (i.e., a signal such as that of Kepler-61 could only have been detected around 63\% of main-sequence Kepler targets).  The corrected planet count is then (22 - 2.09)$\times$1.59 = 31.66.  With this, the planet prior becomes 31.66 / 138,253 = 2.29$\times10^{-4}$. The final odds ratio for Kepler-61b is then 2.29$\times10^{-4}$ / 7.27$\times10^{-9}$ = 31,500, which 1.75 times more significant as we found with a more simplified approach.

\begin{figure}
\begin{center}
 \includegraphics[height=6in]{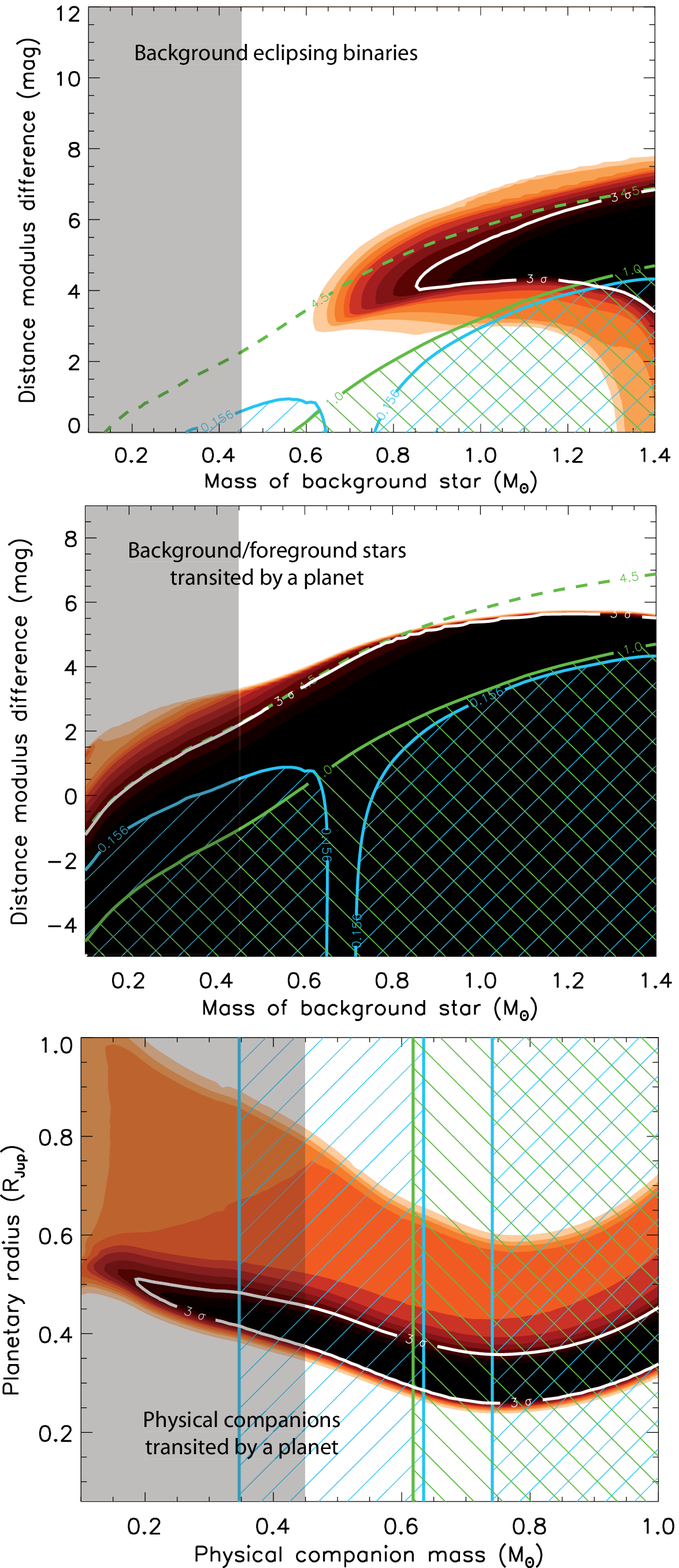} 
 \caption{BLENDER goodness-of-fit contours for Kepler-61b corresponding to the three different scenarios that contribute to the overall blend frequency: background eclipsing binaries (top), background or foreground stars transited by a planet (middle), and physical companions transited by a planet (bottom). Solid colored contours depict the difference in $\chi^{2}$ between an authentic transit model and a blend fit with those parameters. Only blends within the solid white contour acceptably match the \kepler\ light curve (3$\sigma$ difference in $\chi^{2}$ between the blend and transit model fit; see \citealt{Fressin11}, while red, orange, and yellow contours correspond to blend models disfavored by the \kepler\ photometry by 4, 5, and 6$\sigma$, respectively. The axes in each panel represent two of the dimensions of parameter space for blends. For the top two diagrams the vertical axis depicts the distance modulus difference between the two stars, while the horizontal axis corresponds to the mass (spectral type) of the putative secondary star. In the bottom panel (hierarchical triple scenario), the vertical axis corresponds the size of the planet transiting the companion star. The cyan cross-hatched areas indicate regions of parameter space ruled out because the resulting Sloan $r'$-2MASS $K$ color of the blend is either too red (left) or too blue (right) compared to the measured color, by more than 3$\sigma$ (0.15 mag). The green hatched regions indicate blends that are ruled out because the additional star is less than 1 magnitude fainter than the target and would have been detected spectroscopically. Finally, the gray areas on the left represents the constraint from our \spitzer\ observations. The diagonal dashed green lines in the top two panels indicate the faintest blends that can mimic the transit: approximately $\Delta$Kp = 4.5 mag both for background eclipsing binaries and for background/foreground stars transited by a planet.}
  \label{fig:blender}
\end{center}
\end{figure}

Blends that include a companion star $<$0.45 $M_{\odot}$ would produce transit depths inconsistent with our \spitzer\ observations, since they would produce transits depths more than 3$\sigma$ deeper than we measure at 4.5 $\mu$m. These blends are thus excluded. For Kepler-61, the \spitzer\ results exclude all remaining possible physically bound configurations, which would have been the major cause of false positives otherwise. We depict this constraint in the bottom panel of Figure \ref{fig:blender}. 

We note that we have assumed that any signal with a SNR larger than 7.1 would have been recovered by the \kepler\ pipeline as a KOI, to compute the incompleteness correction factor. This optimistic hypothesis is a conservative one in our case, since a more realistic detection model would further increase the incompleteness correction to our planet prior. We conclude that Kepler-61 is an authentic 2.15 $R_{\oplus}$ planet with a high degree of confidence.  We depict an illustration of the \blender\ constraints on false positives for Kepler-61 in Figure \ref{fig:blender}. 

\section{Discussion and Conclusions}

\subsection{Transit Times}

\begin{figure}[h!]
\begin{center}
 \includegraphics[width=5in]{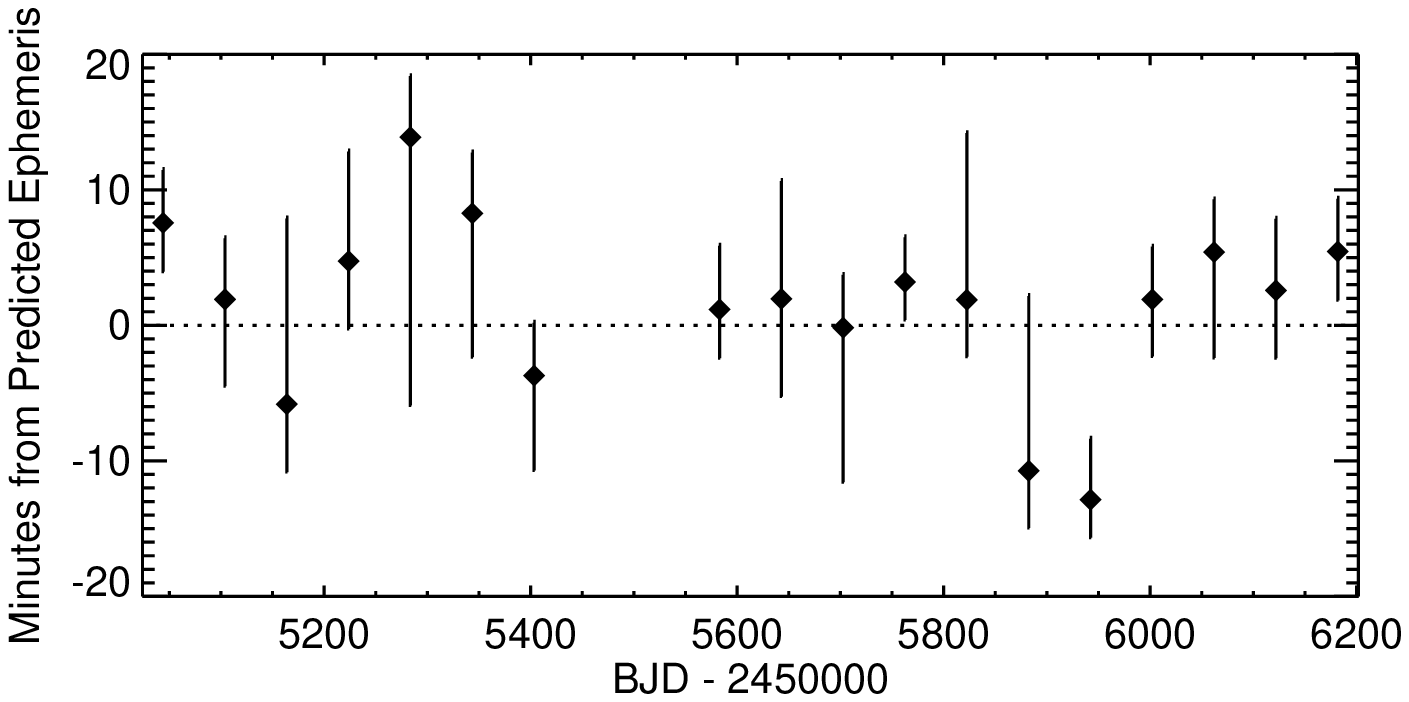} 
 \caption{\kepler\ transit times for Kepler-61 from Quarters 1-14, as compared to the best linear ephemeris model.} 
  \label{fig:times_5}
\end{center}
\end{figure}

We depict the transit times of Kepler-61b in Figure \ref{fig:times_5}. We report no significant deviation from a linear ephemeris. 

\subsection{Theoretical Composition of Kepler-61b}
\subsubsection{Bulk Composition and Atmosphere}
While we cannot estimate the mean density of Kepler-61b without a measurement of its mass, we can still discuss plausible compositions, given its equilibrium temperature and radius. There now exist a sizable set of exoplanets with radii in the 1.0--3.0 $R_{\oplus}$ range with dynamically measured masses, though these span a large range of bulk densities from 0.7 g cm$^{-3}$ in the case of Kepler-11f to 10.4 g cm$^{-3}$ in the case of CoRoT-7b (Kepler-68c may comprise a very dense exception, with $\rho=28^{+13}_{-23}$, but the density range is large and relatively unconstraining). We list published masses, radii, and mean densities from the literature in Table 4. The two planets nearest to Kepler-61 in radius are 55 Cancri e \citep{Winn11} and Kepler-68b \citep{Gilliland13}, the radii of which lie within 0.15 $R_{\oplus}$ of the radius of Kepler-61. \cite{Carter12} found that, even within the small known sample of super-Earths with measured radii, a trend is apparent for planets with equilibrium temperatures $<1200$ K: these tend to have ``mini-Neptune'' compositions, with mean density $<$3.5 g cm$^{-3}$. However, these planets (Kepler-11b, d, e, f, \& g, described by \citealt{Lissauer11}, and GJ~1214b, described by \citealt{Charbonneau09}) are also near to or larger than 2 $R_{\oplus}$, whereas all planets with measured densities higher than approximately 7 g cm$^{-3}$ have radii smaller than 2 $R_{\oplus}$.  It is therefore unclear whether the low density of this set of cooler planets is attributable to their planetary radius or their insolation, or both; this question is explored in greater detail in \cite{Weiss13}, who incorporated both quantities in relation to planetary mass in their exoplanetary mass-radius relation. 

We also consider the theoretical atmospheric content of a 2.15 $R_{\oplus}$ planet. \cite{Rogers11} consider two scenarios (core accretion and outgassing) by which planets in the 2--4 $R_{\oplus}$ regime might retain a substantial hydrogen and helium envelope. Though that work focuses on temperatures $>$ 500 K, a cooler temperature would extend still longer the lifetime of a putative hydrogen/helium envelope. For example, if Kepler-61 formed by core-nucleated accretion beyond the snow line, at 500 K (substantially warmer), a hydrogen helium envelope fraction of 0.1\% by mass is plausible for timescales $<$1 Gyr. Alternatively, it the hydrogen content of the atmosphere is outgassed from the planet (assumed to be formed from iron enstatite), a mass fraction of 1\% by mass is plausible for timescales as long as 100 Gyr. These timescales (and their corresponding atmospheric mass fractions) should be considered lower bounds, given these formation scenarios, since a cooler planet like Kepler-61 will retain an atmosphere for a longer duration, all else being equal. The synthetic planetary radius distribution generated by the formation models of \cite{Mordasini12}, which assume a primordial hydrogen/helium envelope, furnishes a good match to the \kepler\ candidates for planets larger than 2 $R_{\oplus}$, but diverges from the \kepler\ results for smaller radii. This result may be attributable to the more terrestrial composition of planets smaller than 2 $R_{\oplus}$, for which the assumption of a hydrogen/helium envelope is no longer valid \citep{Mordasini12}.

We conclude that a density larger than 7 g cm$^{-3}$, which has only be observed for planets $<$2 $R_{\oplus}$, is unlikely for Kepler-61. Given its radius and comparatively low stellar insolation, its mass may be closer to the 3-6 g cm$^{-3}$ density range bracketed by 55 Cnc e or Kepler-68b with similar radii. If we apply the relation derived by \cite{Weiss13} from the sample of exoplanets with radius measurements and masses $<$150$M_{\oplus}$ (another power law applies for more massive planets), we find a predicted mass and density for Kepler-61b of 3.2 $M_{\oplus}$ and 2.4 cm$^{-3}$, respectively, near to that of Kepler-11b \citep{Lissauer11}. 

\begin{deluxetable*}{rrrrr}
\tabletypesize{\scriptsize}
\singlespace
\tablecaption{Properties of Transiting Planets from 1.4--3.0 $R_{\oplus}$ with Dynamically-Measured Masses}
\label{tbl:superearths}
\tablewidth{0pt}
\tablehead{
\colhead{Name} & \colhead{Radius} & \colhead{Mass} & \colhead{Mean Density} & \colhead{Reference}\\
\colhead{ } & \colhead{[$R_{\oplus}$]} & \colhead{[$M_{\oplus}$]} & \colhead{[g cm$^{-3}$]} & \colhead{ }\\}
\startdata
Kepler-68c & 0.953$^{+0.037}_{-0.042}$ & 4.8 $^{+2.5}_{-3.6}$ & 28$^{+13}_{-23}$ & \cite{Gilliland13} \\
Kepler-10b & 1.416$^{+0.033}_{-0.036}$ & 4.56$^{+1.17}_{-1.29}$ & 8.8$^{+2.1}_{-2.9}$ & \cite{Batalha11} \\
Kepler-36b &1.486$\pm$0.035 & 4.45$^{+0.33}_{-0.27}$ & 7.46$^{+0.74}_{-0.59}$ & \cite{Carter12} \\
CoRoT-7b & 1.58$\pm$0.10 & 7.42$\pm$1.21 & 10.4$\pm$1.8\tablenotemark{a} & \cite{Bruntt10}, \\
  &  &  &  &  \cite{Hatzes11} \\
Kepler-20b & 1.91$^{+0.12}_{-0.21}$  & 8.7$\pm$2.2  & 6.9$^{+5.3}_{-2.6}$\tablenotemark{b}  & \cite{Gautier12} \\
Kepler-11b &  1.97$\pm$0.19 & 4.3$^{+2.2}_{-2.0}$  & 3.1$^{+2.1}_{-1.5}$  & \cite{Lissauer11} \\
Kepler-18b & 2.00$\pm$0.10  & 6.9$\pm$3.4 & 4.9$\pm$2.4 &  \cite{Cochran11}\\
55 Cnc e & 2.00$\pm$0.14  & 8.63$\pm$0.35  & 5.9$_{-1.1}^{+1.5}$  & \cite{Winn11} \\
Kepler-68b & 2.31$\pm$0.07 & 8.3$\pm$2.3 &  3.32$\pm$0.92 & \cite{Gilliland13} \\
Kepler-11f & 2.61$\pm$0.25  & 2.3$^{+2.2}_{-1.2}$  & 0.7$^{+0.7}_{-0.4}$  & \cite{Lissauer11} \\
GJ 1214b &  2.678$\pm$0.13 &  6.55$\pm$0.98 &  1.87$\pm$0.4 & \cite{Charbonneau09} \\
\enddata
\tablenotetext{a}{Differing mass measurements of CoRoT-7b \citep{Queloz09, Pont11} furnish different mean densities; we have stated the most recently published values.}
\tablenotetext{b}{Mean density calculated from stated 1$\sigma$ limits in radius and mass.}
\end{deluxetable*}

There also exist theoretical constraints on the sustainability of super-Earth atmospheres for higher mean molecular weights.  In particular, \cite{Heng12} consider the stability of high mean molecular weight atmospheres belonging to super Earths orbiting low-mass stars in particular. The proximity of the habitable zone to the star means that many super Earths will be spin-synchronized, with a permanent day and night side. In particular, the timescale for spin synchronization is given by \cite{Bodenheimer01} and stated in terms of orbital frequency $\Omega$ by \cite{Heng12}:

\begin{equation}
t_{syn}=\frac{8Q}{45\Omega}\left(\frac{\omega}{\Omega}\right)\left(\frac{M_{p}}{M_{\star}}\right)\left(\frac{a}{R_{p}}\right)^{3}
\end{equation}

\noindent where the planet's initial rotational frequency is given by $w$, $Q$ is the tidal quality factor (and is believed to lie within the range of 10--100 for rocky exoplanets, and in the $10^{5}$--$10^{6}$ range for gas giants, as stated in \citealt{Goldreich66}). Even with extremely rapid initial rotational periods of the planet (e.g. 0.1 day) and values for $Q$ which approach that of gas giants, Kepler-61 is close enough to its host star where the spin synchronization timescale is less than 1 Myr. 

This poses a problem for atmospheric stability unless the zonal winds' ability to redistribute heat in the atmosphere outstrips the radiative timescale of the atmosphere. If this condition (namely, that the advective timescale is shorter than the radiative timescale) does not hold, then the low temperature of  ``night side'' of the planet can allow heavier elements to condense out, leaving the atmosphere unstable. An atmosphere comprising heavier elements has a longer advective timescale, since the wind speed is slowed as mean molecular weight increases (similarly to the sound speed). For this reason, Earth-like atmospheres (with mean molecular weights of 30) are particularly susceptible to instability. The fact that Kepler-61 orbits a late K dwarf translates to a radiative longer than the advective timescale \citep{Heng12}, so an Earth-like atmosphere would remain stable. For later M stars (for example, an M 3.5V star, as adopted as a trial case by \citealt{Heng12}), the radiative timescale at 0.25 AU is shorter, so a 2.5 $R_{\oplus}$ planet possessing an atmosphere with mean molecular weight of 30 would be potentially unstable.

\subsection{Future Prospects}
We comment briefly on the feasibility of atmospheric characterization of Kepler-61b. It orbits a small star and may possess a hydrogen and helium atmosphere, both of which are favorable circumstances for transmission spectroscopy. To perform a basic estimation of the expected change in transit depth at an optically thick wavelength, we consider the atmosphere to be a ring with scale height $H$, where $H=kT_{p}/\mu_{m} g$ (and k is Boltzmann's constant, $T_{p}$ is the temperature of the planet, $\mu_{m}$ is the mean molecular weight of the atmosphere, and $g$ is the surface gravity of the planet). If we use a mass estimate of 8 $M_{\oplus}$ for Kepler-61b (near that measured by \citealt{Gilliland13} for Kepler-68b, with a similar radius), then we expect a surface gravity of 17 m s$^{-2}$. If we assume the most optimistic case from a detectability standpoint, we also use molecular weight of 2 (corresponding to the hydrogen-rich scenario). Employing the the same equation to estimate the change in transit depth attributable to the atmosphere as \cite{Millerricci09}, we expect a change in transit depth given by:

\begin{equation}
\Delta D \approx \frac{2\pi R_{p}H}{\pi R_{\star}^2}=\frac{2R_{p}H}{R_{\star}^2}
\end{equation}

\noindent which equates to $\Delta D$=10 ppm, if we employ the values for $R_{p}$ and $R_{\star}$ given in Table 2. If we instead assume a mass of 2.3 $M_{\oplus}$ , like that of the 2.6 $R_{\oplus}$ planet Kepler-11f \citep{Lissauer11},  then $\Delta D$ is correspondingly three times larger, at 30 ppm. This signal is approximately one-tenth the size of the 0.5 mmag values which might have been detectable in the atmosphere of GJ 1214b by \cite{Berta12} using the Wide Field Camera 3 on board the {\it Hubble} Space Telescope. However, Kepler-61 is also 25 times dimmer in $K$ band than GJ~1214, rendering the detection of an atmosphere around Kepler-61 out of the reach of current instruments.

Similarly, the radial velocity amplitude of Kepler-61b is increased by the small mass of the host star. In this case, assuming again a mass of 8 $M_{\oplus}$ for the planet and a mass of 0.64 $M_{\odot}$ for the star, the planet induces a 1.8 m s$^{-1}$ motion of its star. However, though measuring a radial velocity signature of several meters-per-second has been achieved for dozens of exoplanets, these are all around very nearby stars. Kepler-61b, with its Kepler magnitude Kp of 15.0, is probably too dim for such study with current instruments. However, gathering additional \kepler\ observations of Kepler-61b will be helpful, particularly given the fact that it will be observed in short cadence mode for Quarter 12 onward. 

\subsection{Conclusions}

We present the validation and characterization of Kepler-61b, a 2.15$\pm$0.13 $R_{\oplus}$ exoplanet with equilibrium temperature of 273$\pm$13 K, orbiting a late K dwarf. We determine that the planetary hypothesis for the transit signature of Kepler-61b is 30,000 times more likely than the false positive hypothesis, folding together evidence from high-resolution imagery, the stellar colors, the centroid position of the star from the \kepler\ images, the depth of the transit in the 4.5 $\mu$m bandpass from \spitzer, and from the detailed comparison of the \kepler\ photometry to theoretical light curves of both planetary transits and stellar blends.  Our measurement of the radius and temperature of the star Kepler-61 is based upon a weighted mean of the directly measured radii and temperatures of a subset of nearby stars with the same spectral type, which quantities we apply as priors in our characterization of the planet. We present $K$--band spectra and newly derived metallicities for this set of four similar stars, as well as for Kepler-61. The application of this empirical method, as compared to characterization from $K$--band spectra and stellar evolutionary models, ultimately increased the size and temperature of the planet by 10\%. We consider plausible compositions for Kepler-61b from the set of planets with similar radii and dynamically measured masses, as well as from mass-radius relationships for exoplanets. We conclude that the planet is likely slightly too large to be terrestrial in composition, and likely possesses a significant atmosphere.

We thank Perry Berlind and Mike Calkins at the Fred Lawrence Whipple Observatory for gathering the FAST spectra of Kepler-61 and GJ~380. We thank Courtney Dressing for applying the methodology of \cite{Dressing13} to deduce the physical properties of Kepler-61 and sharing these values with us. We thank Philip Muirhead, Katherine Hamren, Everett Schlawin, B\'arbara Rojas-Ayala, Kevin Covey, and James Lloyd for gathering, reducing, and sharing the TripleSpec $K$-band spectrum of Kepler-61. We thank the \spitzer\ team at the Infrared Processing and Analysis Center in Pasadena, California, and in particular Nancy Silbermann for scheduling the \spitzer\ observations of this program. This work was performed in part under contract with the California Institute of Technology (Caltech) funded by NASA through the Sagan Fellowship Program. It was conducted with observations made with the Spitzer Space Telescope, which is operated by the Jet Propulsion Laboratory, California Institute of Technology under a contract with NASA. Support for this work was provided by NASA through an award issued by JPL/Caltech. This work is also based on observations made with \kepler, which was competitively selected as the tenth Discovery mission. Funding for this mission is provided by NASA's Science Mission Directorate. The authors would like to thank the many people who generously gave so much their time to make this Mission a success.

\newpage


\end{document}